\documentclass{article}

\usepackage{arxiv}
\usepackage[utf8]{inputenc}
\usepackage[T1]{fontenc}
\usepackage[hyphens]{url}
\RequirePackage[colorlinks,citecolor=blue, urlcolor=blue, linkcolor=blue]{hyperref}
\usepackage[titletoc,title]{appendix}
\usepackage[round]{natbib} 
\usepackage[table]{xcolor}
\definecolor{lightgray}{gray}{0.9}
\usepackage{url}
\usepackage{booktabs}
\usepackage{amsfonts}
\usepackage{nicefrac}
\usepackage{microtype}
\usepackage{graphicx}
\usepackage{doi}
\usepackage{amsthm}
\usepackage{amsmath}
\usepackage{amssymb}
\usepackage{mathtools}
\usepackage{chngcntr}
\usepackage{caption}
\usepackage{ragged2e}

\DeclarePairedDelimiter\abs{\lvert}{\rvert}
\DeclarePairedDelimiterX\set[1]\lbrace\rbrace{#1}

\DeclareMathOperator{\sgn}{sgn}
\DeclareMathOperator{\Exp}{\mathbb{E}}

\DeclareMathOperator*{\argmax}{argmax}

\newcolumntype{R}[1]{>{\RaggedLeft\arraybackslash}p{#1}}

\newcommand\nR{31,081}
\newcommand\nT{26,052}
\newcommand\nRwT{1,149}
\newcommand\nRwoT{29,932}

\newcommand\possGivTO{84.9}
\newcommand\possGivNoTO{69.8}

\newcommand\nUnits{4,684}
\newcommand\nTreatments{834}
\newcommand\nControls{3,850}

\newcommand\PCtxtx{0.612}
\newcommand\PCconcon{0.847}
\newcommand\att{$-0.35$}

\newcommand\attPvalue{$<0.001$}

\title{The causal effect of a timeout at stopping an opposing run in the NBA}

\date{July 6, 2021}

\author{ \href{https://orcid.org/0000-0002-8852-5594}{\includegraphics[scale=0.06]{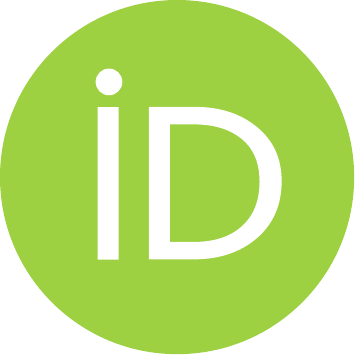}\hspace{1mm}Connor ~Gibbs} \\
	Department of Statistics\\
	Colorado State University\\
	Fort Collins, CO 80523 \\
	\texttt{Connor.Gibbs@colostate.edu} \\
	\And
	\href{https://orcid.org/0000-0002-0092-4532}{\includegraphics[scale=0.06]{orcid.pdf}\hspace{1mm}Ryan ~Elmore} \\
	Department of Business Information and Analytics\\
	University of Denver\\
	Denver, CO 80208 \\
	\texttt{Ryan.Elmore@du.edu} \\
	\AND
	\href{https://orcid.org/0000-0003-3736-2219}{\includegraphics[scale=0.06]{orcid.pdf}\hspace{1mm}Bailey ~Fosdick} \\
	Department of Statistics\\
	Colorado State University\\
	Denver, CO 80523 \\
	\texttt{Bailey.Fosdick@colostate.edu} \\
}

\renewcommand{\shorttitle}{Timeouts in the NBA}

\hypersetup{
pdftitle={The causal effect of a timeout at stopping an opposing run in the NBA},
pdfsubject={stat.AP},
pdfauthor={Connor ~Gibbs, Ryan ~Elmore, Bailey ~Fosdick},
pdfkeywords={causal inference, matching, sports statistics},
}

\begin{document}
\maketitle

\begin{abstract}
In the summer of 2017, the National Basketball Association reduced the number of total timeouts, along with other rule changes, to regulate the flow of the game. With these rule changes, it becomes increasingly important for coaches to effectively manage their timeouts. Understanding the utility of a timeout under various game scenarios, {\em{e.g.}}, during an opposing team's run, is of the utmost importance. There are two schools of thought when the opposition is on a run: (1) call a timeout and allow your team to rest and regroup, or (2) save a timeout and hope your team can make corrections during play. This paper investigates the credence of these tenets using the Rubin causal model framework to quantify the causal effect of a timeout in the presence of an opposing team's run. Too often overlooked, we carefully consider the stable unit-treatment-value assumption (SUTVA) in this context and use SUTVA to motivate our definition of units. To measure the effect of a timeout, we introduce a novel, interpretable outcome based on the score difference to describe broad changes in the scoring dynamics. This outcome is well-suited for situations where the quantity of interest fluctuates frequently, a commonality in many sports analytics applications. We conclude from our analysis that while comebacks frequently occur after a run, it is slightly disadvantageous to call a timeout during a run by the opposing team and further demonstrate that the magnitude of this effect varies by franchise.
\end{abstract}


\section{Introduction} \label{intro}

In game five of the 2019 National Basketball Association (NBA) finals, the Toronto Raptors' Kawhi Leonard scored ten straight points to give the Raptors a 103-97 lead over their opponents, the Golden State Warriors. The Toronto Raptors' coach, Nick Nurse, called a timeout immediately after Kawhi's last basket with three minutes and five seconds left on the clock. Following the timeout, the Toronto Raptors' offense became stagnant, scoring only two points during the remainder of the game, while the Warriors scored nine. The Golden State Warriors won 106 to 105. After the game, Nick Nurse was chastised for his decision to call a timeout \citep{nn1-post, nn2-bi, nn3-usatoday}. ESPN commentator Stephen A. Smith even blamed Nick Nurse for the Raptors' loss, citing his timeout as the disturbance to the Raptors' run, {\em i.e.}, when one team (Raptors) has significantly outscored the other team (Warriors) in a short period of time \citep{stepha}. At the heart of this commentary is the belief that timeouts cause a disruption to a team that is on a run, {\em i.e.}, scoring during a run. If true, this would be valuable information for coaches who must choose when and under what circumstances to call a timeout.

The NBA is pressuring coaches to call fewer timeouts. In the summer of 2017, the NBA reduced the total number of timeouts from 18 to 14, among other rule changes, to regulate the flow of the game \citep{nba-rule-changes, crossover2019}. With fewer timeouts, it becomes increasingly important for coaches to effectively manage their timeouts. Thus, understanding the utility of a timeout under various in-game scenarios, {\em{e.g.}}, during an opposing team’s run, is of the utmost importance. 

Whether to call a timeout during an opposing run is highly debated among professional coaches. There are two schools of thought when the opposition is on a run: (1) call a timeout and allow your team to rest and regroup, or (2) save a timeout and hope your team can make the needed corrections during play. According to \citet{athletic2018}, coach Rick Carlisle of the Indiana Pacers and recently with the Dallas Mavericks is known for calling timeouts in ``an obvious situation, like stopping a run by the opponent ...'' He tends to coach by the first philosophy. On the other hand, coach Mike D'Antoni, most recently of the Houston Rockets, tends to refrain from calling a timeout, citing his trust in his team's ability to ``break runs up with their stellar plays'' \citep{athletic2018}. He tends to coach by the second philosophy. This paper investigates the credence of these two coaching philosophies by estimating the causal impact of a timeout in the presence of an opposing team's run. 

Runs are largely studied within the context of the hot hand phenomenon, or the belief that a player's current shooting success is indicative of their short-term, future shooting success \citep{gilovich1985hot, koehler2003hot, avugos2013hot, miller2018surprised}. The literature surrounding the efficacy of timeouts in the NBA exists but is relatively sparse. \citet{saavedra2012coaching} define a timeout factor to gauge team performance after a timeout relative to their average, allowing the authors to study the relationship between the timeout factor and the scoring dynamics. They found the timeout factor played a minor role in the scoring dynamics. \citet{permutt2011efficacy} studied the efficacy of timeouts at stopping an opposing team's momentum, as defined by six unanswered points. To estimate the effectiveness of timeouts in these situations, the short-term performance of teams when a timeout was called was compared to that when a timeout was not called. This simple comparison fails to account for self-selection bias: bias attributed to the coach's right to choose when to call, or not call, a timeout. 

Technological advances such as player and ball tracking cameras/radar are producing vast quantities of previously unimaginable data. This necessitates that professional teams and researchers must increasingly rely on statistics to better understand the nature of the game.  For example, \citet{franks2015characterizing} apply spatial-temporal methodology to player movement data in order to create advanced defensive metrics in the NBA. \citet{deshpande2020expected} use non-parametric Bayesian analysis and imputation methods to track how completion probability evolves through receivers' routes in the National Football League (NFL). In addition to player movement analyses, \citet{zimmerman2019outline} apply outline analysis to make inference on the geometric attributes of the called strike zone in Major League Baseball (MLB). Furthermore, a larger comparison of team strength within and competitiveness across sports leagues is estimated using Bayesian state-space modeling in \citet{LopezMichaelJ2018Hodt}. In this paper, we leverage the vast play-by-play data provided by the NBA to estimate the causal impact of a timeout, an oft-debated topic.

Experimental studies within sports, however, are largely impossible due to the importance placed on each coaching decision. At the same time, sports fans, commentators, and analysts enjoy questioning the causal impact of such decisions. Examples include questioning the causal effect of going for it on fourth down in the NFL \citep{yam2019}, clearing the puck in the National Hockey League \citep{toumi2019}, or taking a pitch during a 3-0 count in MLB \citep{vock2018estimating}. When experiments are impractical or impossible, causal inference, a field dedicated to estimating causal effects from observational data, is used. In the context of the NBA, \cite{assis2020stop}  estimate the causal effect of a timeout on team performance through scoring dynamics before and after a timeout. They concluded timeouts have no effect on teams' performances, yet they considered all timeouts in a game. We restrict our analysis to include only those timeouts called in response to a run. To our knowledge, no study has explored the effectiveness of timeouts at stopping an opposing run through a causal lens.

This manuscript seeks to estimate the causal effect of a timeout at stopping an opposing run. We start by describing the data and formalizing the notion of a run in Section \ref{data}. In Section \ref{methodology}, we review the Rubin causal model and carefully define the units with a discussion of the appropriateness of the stable unit-treatment-value assumption (SUTVA) in this context.  We then describe the matching framework and results before introducing a novel outcome metric. Finally, in Section \ref{results}, we discuss the causal estimate, as well as provide estimates for the causal effect by franchise. In Section \ref{sec:sensitivity}, we investigate the sensitivity of results to alternative run definitions and to unmeasured confounders. We conclude with a discussion in Section \ref{discussion}. 


\section{Data and Notation} 
\label{data}

\subsection{Data}

 The NBA provides access to a wealth of information related to the performance of teams and individuals, as well as game-specific data, through the league's website \citep{nba-site}. A popular R \citep{r-software} package, \texttt{nbastatr} \citep{nbastatr}, allows programmatic access to the NBA's data via the league's Application Programming Interface (API). Using these tools, we acquired play-by-play data for each regular season game played during the 2017-2018 and 2018-2019 seasons. There are 30 teams in the NBA and every team plays 82 regular season games for a total of 1,230 unique games. All the events that occur during a game are recorded and are exposed through the API. In total, the entire play-by-play data set from the 2017-18 and 2018-19 seasons consists of 1,144,461 events, each described by 18 variables. On average, there are approximately 465 time-ordered events per game (standard deviation $33.9$).

An \textit{event} in the play-by-play data set is any time-stamped action that is recorded by the scorekeeper during the game. For example, events may include a made (or missed) field goal, a rebound after a missed field goal (offensive or defensive), a foul, a timeout, among others. Specific events from the Houston Rockets at Denver Nuggets game on February 1, 2019 include James Harden making a three point shot approximately six minutes into the first period ``Harden 25' 3PT Jump Shot (10 PTS)'', Nikola Jokic grabbing a defensive rebound with three minutes left in the fourth period ``Jokic REBOUND (Off:4 Def:0),'' and as is of key importance to this work, the Nuggets calling a fourth-period timeout immediately after another James Harden three pointer ``NUGGETS Timeout: Regular (Full 5 Short 0).'' 

In this analysis, we term a \textit{play} to consist of a set of simultaneous events, as tagged by the play clock. For example, when a foul is committed, two free throws may follow. At the time of the foul, the clock is stopped, and the fouled player attempts his two free throws. The foul and the subsequent free throws occur at the same point in time in the game since the clock is stopped. The foul along with the two free throws are three events which make up one play, according to our definitions.

We simplify the data set by recording each timeout and reducing each play to be represented by a single event. Specifically, we retain the last recorded scoring event in the play, unless the play contained no scoring events, in which case we retain the last event in the play. In this process of removing unnecessary events, we create an indicator variable associated with each play to document whether a timeout was called. In total, there were 778,828 plays in the 2017-18 and 2018-19 NBA seasons.

\subsection{Notation}

The principle question in this investigation is whether or not a timeout has an effect on a game following a ``run,'' where a run, colloquially, is when one team significantly outscores the other in a short amount of time. Particularly, we are interested in the timeout's effect on the team that is not ``on the run'' ({\emph{i.e.,}} not scoring during the run). To address this question, we first formally define a run in the context of the data available through the NBA's API.

 Let $t$ represent the time in minutes in an NBA game, $t\in[0, 48]$, and denote the home team score and the away team score as $h(t)$ and $a(t)$, respectively. Note that we ignore over-time in this analysis. Define the \textit{score difference} at time $t$, $\Delta(t)$, as the home score minus the away score: $\Delta(t) = h(t) - a(t)$. 

Critical to our subsequent development, we define a \textit{run} at time $t$ to be a change in the score difference of at least nine points within the prior two minutes of game time, formally called the \textit{pre-treatment window}. Therefore, we characterize a run by the change in the score difference (\textit{run point total}) and the time required to realize that change (\textit{run duration}). 

Formally, we define the run duration at time $t$, denoted $\delta_{t}$, as the shortest amount of time taken to attain the greatest net change in the score difference to time $t$ in the two minutes prior to time $t$:
\begin{equation} 
\label{eqn:run-duration}
\delta_{t} = \min \left\{\argmax_{d}\Big(\abs{\Delta(t) - \Delta(t-d)} : 0 < d \leq 2\Big) \right\}.
\end{equation}
The \textit{signed run point total} at time $t$, $s(t)$, is taken to be the most extreme net change in the score difference if a run has occurred. That is,
\begin{equation} 
\label{eqn:signed-run}
 s(t) = \begin{cases} 
      \Delta(t) - \Delta(t-\delta_{t}), &\abs{\Delta(t) - \Delta(t-\delta_{t})} \geq 9 \\
      \text{NA}, & \text{otherwise} \\
   \end{cases}.
\end{equation}
If $s(t) > 0$, then the home team is on the run at time $t$, and conversely, the away team is on the run at time $t$ if $s(t) < 0$. The run point total, $r(t)$, is subsequently defined as the magnitude of the signed run point total, $r(t) = \abs{s(t)}$. The sign of a signed run point total lends clarity to which team is on the run at time $t$, a necessary component in defining an interpretable outcome. Of the \nT{} plays involving a timeout, \nRwT{} of them were identified as runs.

\begin{figure}[hbtp!]
    \centering
    \includegraphics[scale=0.62]{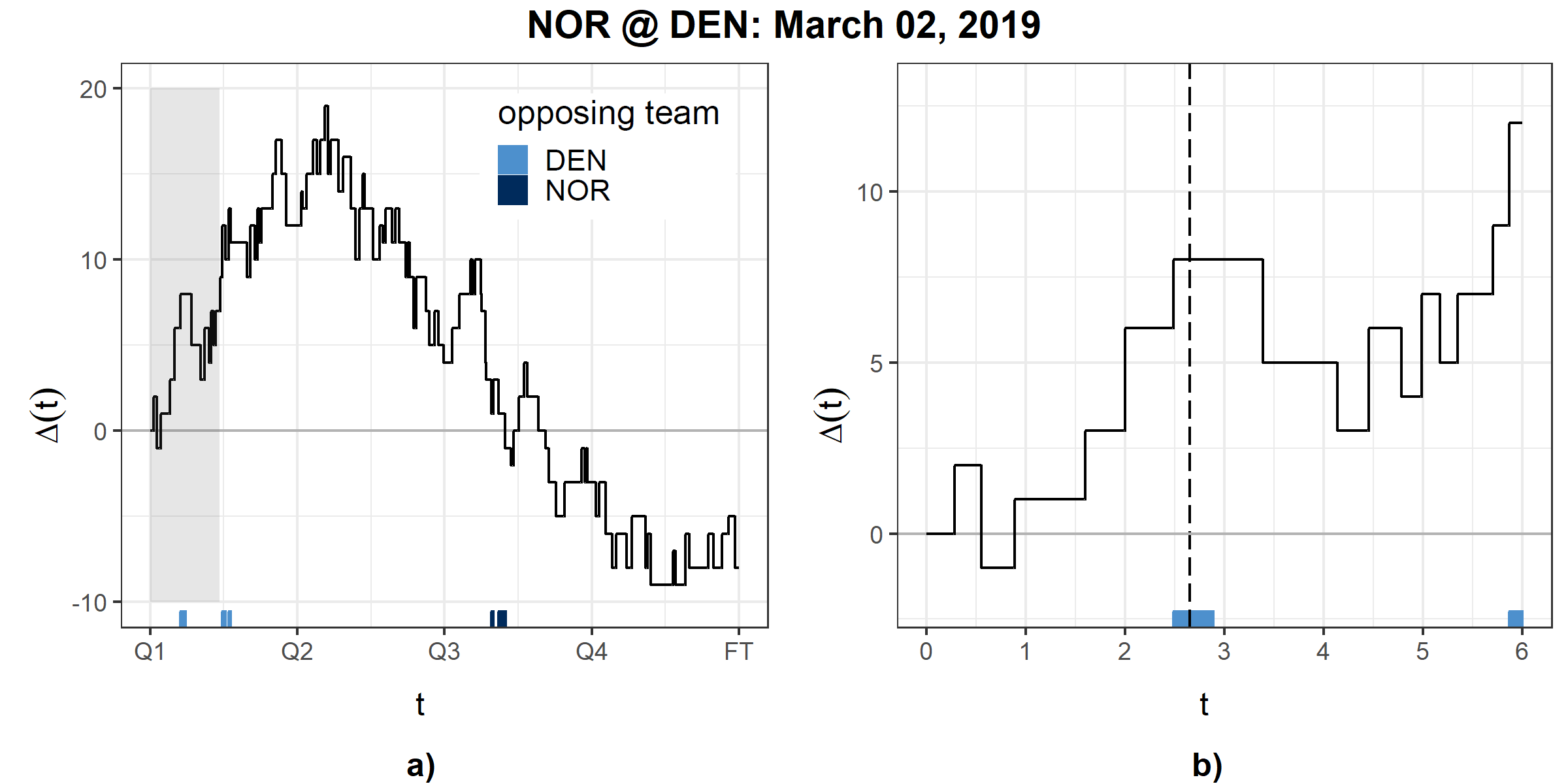}
    \caption{Panel (a) shows the score difference over the course of the game, and panel (b) zooms in to focus on the first six minutes of the game. The horizontal bars (rug) at the bottom of each plot indicate intervals of time when the run criteria are satisfied, colored by the team on the run (opposing team). If the score difference has a positive trajectory immediately prior to the rug, then the home team (Nuggets) was the opposing team, and if it has a negative trajectory, the away team (Pelicans) was the opposing team. The vertical line at $2.65$ minutes in the panel (b) denotes a timeout by the Pelicans, when the run point total, $r(2.65)$, was nine points and the run duration, $\delta_{2.65}$, was $1.76$ minutes.}
    \label{fig:runplot}
\end{figure}

To illustrate the concepts defined above, consider the March 2, 2019 game between the New Orleans Pelicans and the Denver Nuggets, played in Denver, Colorado. The score difference for this game is given in panel (a) of Figure \ref{fig:runplot} and points in time meeting the criteria of a run are indicated in the rug of the plot, colored by the team on the run.

As expected, the majority of plays throughout the game are not runs, and hence the signed run point total, $s(t)$, is NA during that time. During the first period of this game, there are three time intervals when the run criteria are met and two such time intervals in the third period. For the purposes of this analysis, we define the \textit{opposing team} as the team on the run at time $t$ and the \textit{BiT (big trouble) team} as the team which may seek to benefit from a timeout at time $t$.

Consider the first run of the game, highlighted in panel (b) of Figure \ref{fig:runplot}. At roughly two and a half minutes into the game, the Nuggets went on a nine point run in 1.76 minutes. Therefore, at $t=2.65$, the run duration is $\delta_{2.65}=1.76$, where the signed run point total is $s(2.65) = 9$, implying a run point total $r(2.65) = 9$ (see Figure \ref{fig:pre-post-windows}). At this time, the Pelicans may (or may not) choose to call a timeout in an effort to thwart the Nuggets' run, necessarily defining the Nuggets as the opposing team and the Pelicans as the BiT team. 

To further clarify the definition of the run duration in Equation \eqref{eqn:run-duration}, consider panel (a) of Figure \ref{fig:pre-post-windows} where the greatest net change in the score difference to time $t=2.65$ is attained on the interval between $1.76$ and $2$ minutes prior to $t=2.65$ when the score remains constant. Any point in time within this interval would satisfy the requirements of $d$ in Equation \eqref{eqn:run-duration}; however, the outer minimum operation ensures we consider the least amount of time it took the Denver Nuggets to achieve a nine point swing in the score difference, the greatest net change in the score difference to time $t=2.65$ in the pre-treatment window.

Of the 778,828 plays discussed above, \nR{} are classified as a run play ({\emph{i.e.,}} play occurring when the run criteria are satisfied) and \nT{} involved a timeout. There were \nRwT{} run plays with a timeout and \nRwoT{} run plays without a timeout among the \nR{} run plays.

\section{Methodology} \label{methodology}

\subsection{Rubin Causal Model}
\label{sec:rcm}

To isolate the causal effect of a timeout, we would ideally perform a randomized experiment.  In the context of our problem, this could be as easy as flipping a coin on the bench to determine whether or not a team should call a timeout during an opposing team's run.  However, this is obviously impractical due to the importance of each NBA game and the emphasis placed on each coaching decision therein. Thus, we consider how we can leverage the existing observational data to isolate the causal effect of a timeout.

The obvious concern when using observational play-by-play data is that game situations when timeouts are called are likely fundamentally different than game situations when timeouts are not called, bias attributed to the coach's right to choose when to call a timeout.  To tackle this type of problem in general, Rubin's causal inference framework \citep{rubincausalmodel, rubin1976inference, rubin1977assignment} aims to restructure the data to make it most similar to that which might have been observed from a randomized experiment.  This restructuring attempts to remove discrepancies between the distributions of covariates of the timeout (``treated'') and no timeout (``control'') groups, commonly referred to as covariate imbalance.  For example, suppose we identify two instances when an opposing team was on a run: one where a timeout was called and one where a timeout was not called.  To fairly compare the subsequent impacts on the game based on these actions, we must account for which team is in possession of the basketball.  Unsurprisingly, in instances when a team called a timeout, that team had possession \possGivTO{}\% of the time, whereas for plays when a team did not call a timeout, that team had possession only \possGivNoTO{}\% of the time.  This is to be expected as a team can only call a timeout when they have possession, or there is a break between plays, {\em i.e.}, a dead ball.  However, it is important to consider this covariate, among others, when comparing subsequent changes in the game score as it is far less likely for the score to change in a team's favor in the minute immediately following the timeout, or no timeout, if that team does not initially have the ball.

We consider a matching approach to address the covariate imbalance between runs that included a timeout and runs that did not include a timeout and, hence, reduce bias in the estimation of the treatment effect \citep{rosenbaum1985constructing, stuart2010matching}. Effectively, for each run with a timeout, henceforth denoted \textit{RwT}, we find a run without a timeout, henceforth denoted \textit{RwoT}, that most closely matches the in-game situation of the RwT. To formally place this effort in the causal inference framework, we briefly review the standard causal modeling notation as it relates to our specific problem. Let $T_{i}$ be a binary treatment indicator variable, equal to 1 if  a timeout is called for the $i^{th}$ unit (defined in Section \ref{methods-treatments-and-controls}) and 0 if no timeout is called. Further, let $Y_{i}$ be the observed outcome for unit $i$ (defined in Section \ref{sec-outcome}) and $X_{i}$ be the set of covariates for unit $i$ (introduced in Section \ref{sec-prop-score}).

The potential outcomes model expresses the observed outcome $Y_i$ as
\begin{align}
Y_{i} &= 
\begin{cases}
  Y_{i}(0), \quad T_{i} = 0 \\
  Y_{i}(1), \quad T_{i} = 1
\end{cases}
\end{align}
where $Y_{i}(1)$ and $Y_{i}(0)$ denote the potential outcome when a timeout and no timeout is called, respectively. If we observed both potential outcomes for all units, we would naturally estimate the average effect of a timeout as the average difference between the outcome with a timeout minus that without, $\Exp[Y_i(1) - Y_i(0)]$.  Unfortunately, the pair $(Y_{i}(0), Y_{i}(1))$ is not directly observable; instead, we observe $(Y_i, T_i)$. Under random treatment assignment, we could estimate the expected outcomes under timeouts and no timeouts with empirical averages.  However, as previously mentioned, there may be clear differences between the situations when a timeout is called and situations when a timeout is not called, since coaches self-select to enter the treated group. We intentionally narrow our focus here to the \textit{average causal treatment effect on the treated (ATT)}, denoted
\begin{align}
    \text{ATT} &= 
    \Exp_{x\mid{T=1}}\Big[ \Exp\left[ Y_{i}(1) - Y_{i}(0) \mid T_{i} = 1, X_{i} = x\right] \Big].
    \label{eq:matchingATT}
\end{align}
This is the estimand for the average treatment effect for those plays where a coach chose to call a timeout and motivates the matching framework.

In order for ATT to be estimable, strong ignorability must hold \citep{heckman1997matching, heckman1998matching, smith2005does}. This requires two assumptions: conditional independence and positivity. Further, the applicability of the Rubin causal model is founded on the stable unit-treatment-value assumption (SUTVA), which states that (1) there are no hidden variations of treatments ({\em i.e.,} only one form of a timeout), and (2) there is no interference among the units ({\em i.e.,} timeout applied to one unit does not affect the outcome for another unit) \citep{imbens2015causal}. We explore these assumptions as they relate to our data in subsequent sections.



\subsection{Treatments and Controls: Defining the Units} \label{methods-treatments-and-controls}
There are four criteria required of the $j^{th}$ play occurring at time $t_{j}$ to be considered a unit in our sample: (1) a team is on a run, (2) there is no timeout in the pre-treatment window, (3) there is no timeout in the post-treatment window, and (4) the pre-treatment and post-treatment windows are not truncated by the end of the period.  

\begin{figure}[hbtp!]
    \centering
    \includegraphics[scale=0.62]{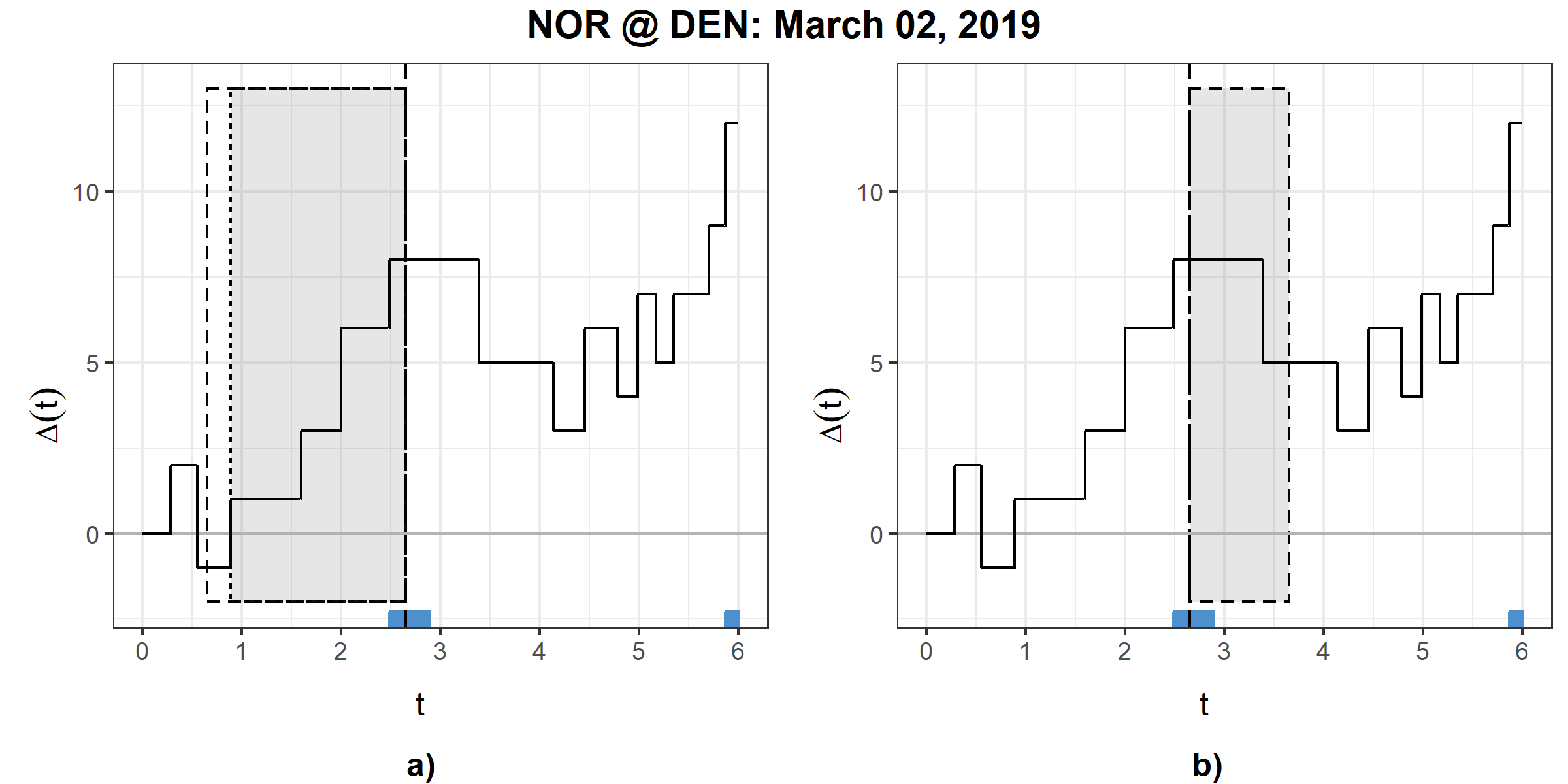}
    \caption{In panel (a), the larger dashed box indicates play occurring in the pre-treatment window of the play of interest denoted by the dashed, vertical line. The smaller shaded box marks the shortest interval of time contained within the pre-treatment window which captures the most extreme net change in the score difference up to the play of interest. If the change in the score difference is greater than 9 in absolute value, then the play of interest is considered a run, the magnitude of the score change is the run point total, and the length of the interval marked by the shaded box is the run duration. Any play occurring in the rug of the plot is considered a run play, colored by the opposing team: the team on the run. In panel (b), the shaded box indicates play occurring in the post-treatment window of the play of interest. The post-treatment window is the interval of time used to compute the outcome of the intervention occurring at the time of the play of interest.}
    \label{fig:pre-post-windows}
\end{figure}

We define the \textit{pre-treatment window} for the $j^{th}$ play as the two-minute interval prior to time $t_j$ (see panel (a) of Figure \ref{fig:pre-post-windows}) and the \textit{post-treatment window} as the one-minute interval of time following $t_j$ (see panel (b) of Figure \ref{fig:pre-post-windows}). Using the pre-treatment window, we assess whether a run has occurred at time $t_j$, while the post-treatment window is used to measure the impact of the intervention at time $t_j$ (see Section \ref{sec-outcome}).  When one of these intervals contains the beginning or end of a period, the interval is said to be truncated, and the play is disregarded.  If a play meets criteria (1)-(4), the play is included in our sample and deemed a treatment unit if a timeout is called at $t_{j}$ and a control unit if no timeout was called at $t_j$.

Therefore, associated with each unit is a three-minute interval of game time used to assess whether a run occurred at $t_j$, characterize whether or not a timeout was called at $t_j$, and measure the effect of the intervention at $t_j$. When these time intervals overlap, we need to focus on potential violations of the SUTVA. As such, criteria (2) and (3) in the unit definition above are aimed at mitigating serious violations while maintaining a reasonable control pool for matching on time-dependent covariates.  

In the following two subsections, we describe all possible violations of SUTVA in the context of our problem, and discuss which are resolved and unresolved by criteria (2) and (3). In discussing these criteria in detail, consider a time $t$ when the run criteria are met. If a timeout is called at time $t$, this play is potentially a treatment and if no timeout is called at time $t$, this play is potentially a control. However, there are a number of reasons why this play might be eliminated from consideration for our study.

\subsubsection{Criterion 2: Timeout in the pre-treatment window}

The pre-treatment window is used to assess whether the play at $t$ is considered a run play or not. The existence of a timeout in the pre-treatment window suggests heterogeneous versions of the treatment. For example, if there was no timeout at $t$, then the play would be under consideration as a potential control. However, the play should not seriously be considered a control unit because the coach recently conferred with his team. Doing so would violate the assumption of no hidden variation among control units. Furthermore, if there was a timeout at $t$, then the coach will have talked to his team at least twice in a short time window. It is fathomable having back-to-back timeouts would be more effective at stopping a run, and we would expect different outcomes than that with a single timeout. This too would be a violation of the assumption of no hidden variation among treated units. In either case, the play at time $t$ is removed to preserve the assumption of no hidden variation.

\subsubsection{Criterion 3: Timeout in the post-treatment window}

The post-treatment window is used to measure the impact of the treatment at time $t$. The existence of a timeout during this time obfuscates the effect of the intervention at $t$.  We interpret this as a second intervention.  For example, suppose the BiT team scores six points quickly after time $t$ and the opposing team calls a timeout. In this case, the outcome for the unit at time $t$ is possibly truncated by the subsequent intervention. This would be a direct violation of the assumption of no interference if the latter timeout is a unit in the study. As such, the play at time $t$ is removed from consideration.

 { 
\renewcommand{\arraystretch}{1.13}
\begin{table}[htpb!]
    \caption{List of covariates with descriptions. The Las Vegas spread, over-under, and moneyline are proxies for the teams' comparative skill and offensive/defensive abilities. The distributions of each covariate by treatment group are provided in Section S5 of the \protect\hyperlink{suppA}{Supplement}.} 
\begin{center}
\rowcolors{1}{}{lightgray}
\begin{tabular}{p{3.5cm}|p{8.5cm}}
  \hline
  Covariates & Description \\ 
  \hline
  Big Trouble (BiT) Team & The team which may seek to benefit from a timeout during an opposing run. \\
  Opposing Team & The team on the run, heavily outscoring the BiT team in a short amount of time. \\
  Run Point Total & The magnitude of the most extreme change in the score difference in the pre-treatment window. This is denoted $r(t)$. \\
  Run Duration & The shortest amount of time taken to attain the most extreme change in the score difference to time $t$. This is denoted  $\delta_{t}$. Equivalently, this is the shortest amount of time taken to attain the run point total. \\
  Time Left & The amount of time left in the game (in minutes). This is equivalent to $48-t$. \\
  Win Probability & The BiT team's probability of winning the game at the time of treatment. \\
  Signed Score Difference (SSD) at Beginning of Run (BOR) & The score difference (expressed as the BiT team score minus opposing team score) when the run began. Positive (negative) values indicate the BiT (opposing) team was leading when the run began. This is equivalent to \nolinebreak ${-\sgn(s(t))\Delta(t-\delta_{t})}$. \\ 
  Signed Score Difference (SSD) at End of Run (EOR) & The score difference (expressed as the BiT team score minus opposing team score) at the time of treatment. Positive (negative) values indicate the BiT (opposing) team was leading at the time of treatment. This is equivalent to $-\sgn(s(t))\Delta(t)$. \\ 
  Possession Indicator & Indicator for ball possession at the time of treatment, equal to 1 if the BiT team has possession and 0 otherwise. \\
  Home Indicator & Indicator for home court advantage, equal to 1 if the BiT team is home and 0 otherwise. \\
  Week in Season & The week in the season. \\
  Over/Under & The Las Vegas over-under prior to the game. \\
  Spread & The Las Vegas spread (expressed as the BiT team score minus opposing team score) prior to the game. \\
  Moneyline & The Las Vegas moneyline prior to the game, assuming a negative value if the BiT team is favored and a positive value otherwise.  \\
  \hline
\end{tabular}
\end{center}
\label{covariate-list}
\end{table}
}

\subsubsection{Final Units}
\label{sec:final_units}

After invoking criteria (1)-(4), we examine the distribution of the covariates by treatment group (see Section S5 of the \hyperlink{suppA}{Supplement}). We remove units with a moneyline larger than 2,400 in absolute value to reasonably justify the positivity assumption. After removal, there remain \nUnits{} runs in the sample, \nTreatments{} of which are RwTs and \nControls{} of which are RwoTs, the final units for our study. The Chicago Bulls have the most RwTs in the analysis set with 41, whereas the Oklahoma City Thunder and the San Antonio Spurs are tied for the fewest with 19. The median number of RwTs per franchise is 27.5 with an interquartile range of 8.5. On the other hand, the most and fewest number of RwoTs in the analysis set belong to the Los Angeles Clippers and Denver Nuggets with 257 and 56, respectively. The median number of RwoTs per franchise is 109.5 with an interquartile range of 76.5. The number of units is also broken down by period in the game and presented in Table \ref{tbl:units-by-period}.

\begin{table}[htpb!]
    \caption{Number of runs with a timeout (RwT) and runs without a timeout (RwoT) by period present in the analysis set.
    } 
\begin{center}
\begin{tabular}{lrrrr|>{\columncolor[HTML]{EFEFEF}}r}
  \hline
  & First & Second & Third & Fourth & Total \\ 
  \hline
  Runs with a Timeout & 266 & 195 & 231 & 142 & 834  \\
  Runs without a Timeout & 1,078 & 1,026 & 1,154 & 592 & 3,850 \\
  \hline
  Total & 1,344 & 1,221 & 1385 & 734 & 4,684 \\
\end{tabular}
\end{center}
\label{tbl:units-by-period}
\end{table}

More details regarding the number of observations remaining after each criterion is applied are provided in Section S1 of the \hyperlink{suppA}{Supplement} along with a summary of the data preparation discussed herein. While the criteria account for the major violations to the SUTVA, there remain minor violations that are not addressed by our criteria. A full expose of these situations is given in Appendix \ref{app}.

\subsection{Propensity Score Model}
\label{sec-prop-score}

The strong ignorability assumption must hold in order to estimate the average treatment effect on the treated. Since there are many factors which influence a coach's decision to call a timeout, we utilize a propensity score model to estimate the probability of calling a timeout in each game scenario and use it in matching treated units to control units. After careful consideration of available covariates for a game situation, those listed in Table \ref{covariate-list} are included in the propensity score model as potentially predictive of a coach's decision to call a timeout.

\begin{figure}[hbtp!]
    \centering
    \includegraphics[scale=0.6]{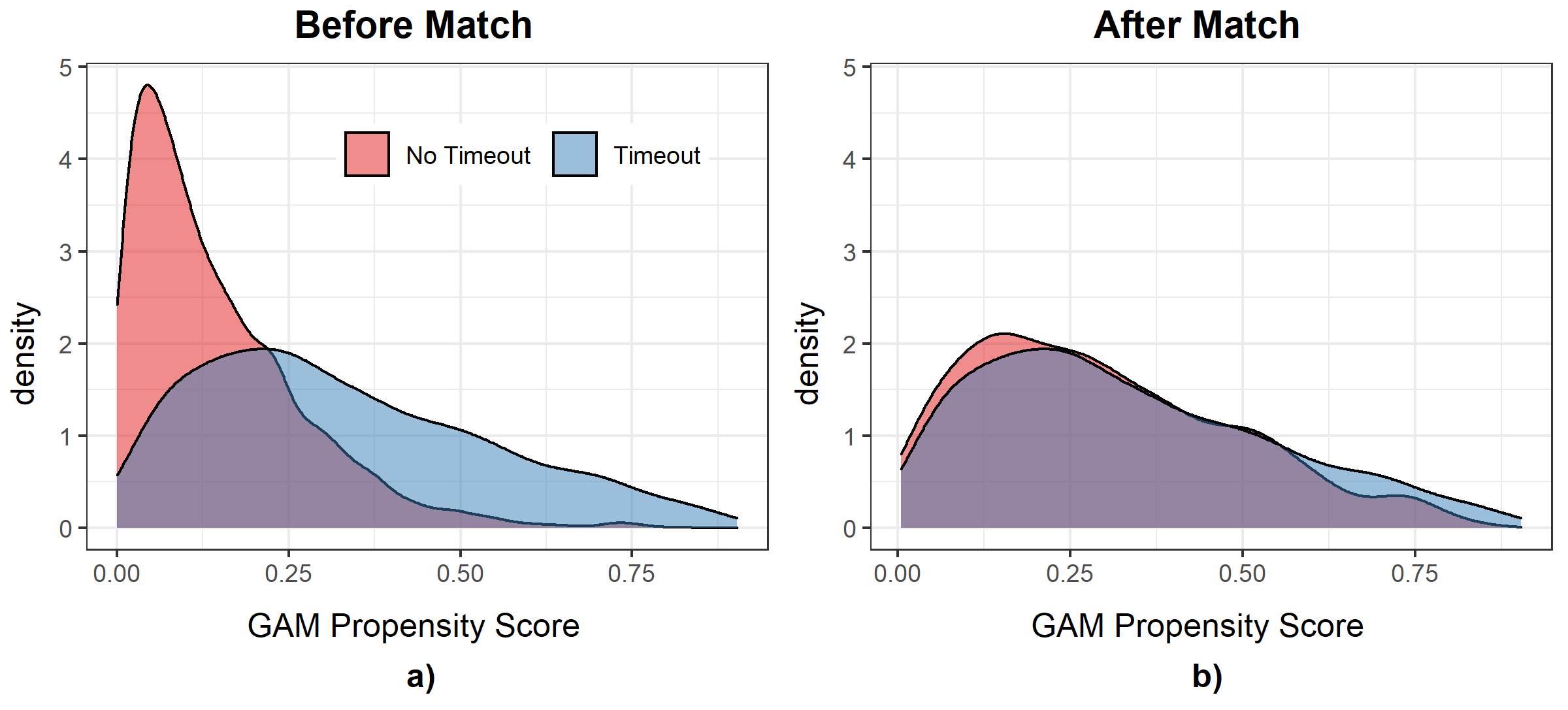}
    \caption{Panel (a) shows that the propensity to call a timeout, estimated using the pre-treatment covariates in Table \ref{covariate-list}, differs between the treated (timeout) and control (no timeout) groups.  This signifies significant covariate imbalance between the groups, which impedes the ability to compare them directly. Panel (b) shows the distribution of propensity scores after matching are quite similar.}
    \label{fig:covbal-propscore}
\end{figure}

We estimate the probability that the BiT team calls a timeout conditioned on the pre-treatment covariates using a generalized additive model (GAM) \citep{gam}. To gauge the predictive validity of the model, we partition the units randomly into a training set (70\%) and a testing set (30\%). We estimate the model using the training set and assess its predictive accuracy on the test set. Averaging over 1,000 Monte Carlo splits, the proportion of classified treatments (controls) that are treatments (controls) is $\PCtxtx{}$ ($\PCconcon{}$), indicating a reasonable model for predicting the treatment group to which each unit belongs. The estimated propensities on the entire set of units are provided in panel (a) of Figure \ref{fig:covbal-propscore} as evidence for covariate imbalance between treated and control units. 

\subsection{Matching}

In seeking to calculate the treatment effect, the naïve approach would be to develop an outcome of interest and calculate the difference in means between the treated units and the control units identified in Section \ref{sec:final_units}. As seen in the panel (a) of Figure \ref{fig:covbal-propscore} and discussed in Section \ref{sec:rcm}, game situations when timeouts are called are fundamentally different from when timeouts are not called. To mitigate the ill effects of self-selection bias, we employ a matching procedure with propensity scores \citep{rosenbaum1983central, stuart2010matching, lopez2017estimation}. 

Two common approaches are generally used to balance covariates: matching on the distance between the covariate vectors, such as Mahalanobis distance, or matching based on the estimated propensity scores. Here we employ a hybrid approach introduced by \cite{genmatchingtheory} that matches based on the distance between the covariate vectors and the propensity scores through minimization of a generalized version of Mahalanobis distance (GMD). In this approach, the GMD has a weight parameter, and the weights of the covariates and propensity score are chosen to minimize the largest individual discrepancy using $p$-values from Kolmogorov-Smirnov tests and paired $t$-tests. This estimation procedure is implemented in the \texttt{Matching} package in \texttt{R} using a genetic search algorithm \citep{genmatchingpackage}. Details regarding the function arguments used can be found in Appendix \ref{app:gen_match}.

In practice, every observation in the treated group is matched to an observation in the group of potential controls in a one-to-many fashion. That is, a potential control can feasibly serve as the control for more than one treatment. As an example, the identified match for the treated unit in Figure \ref{fig:runplot} is a play from February 06, 2019 when the Washington Wizards played the Milwaukee Bucks in Milwaukee. The identified match featured a run of nine points attained in 1.73 minutes. At the time of treatment, there were 38.08 minutes left in the game. In both situations, the BiT team was away and did not have possession at the time of treatment. The probability of calling a timeout (estimated from the GAM) was 0.32 for the RwT versus 0.28 for the matched RwoT. Thus, we see that in both scenarios the covariates were largely comparable. For more detail regarding covariate balance, including the distribution of the covariates by treatment group after matching, see Section S5 of the \hyperlink{suppA}{Supplement}.

\begin{figure}[hbtp!]
    \centering
    \includegraphics[scale=0.8]{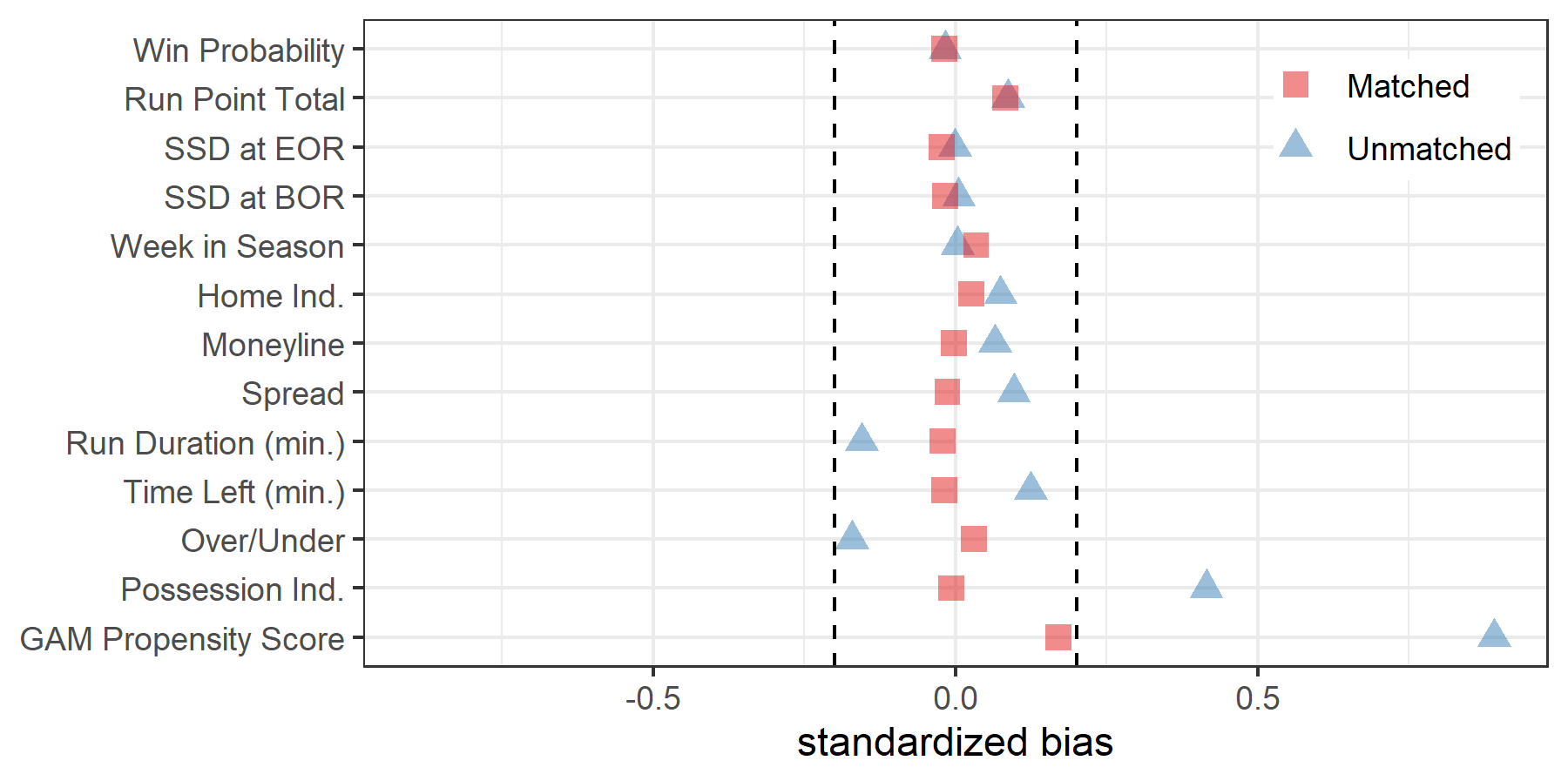}
    \caption{Standardized bias is often used to gauge covariate balance before and after matching or other trimming procedures. The propensity to call a timeout is largely imbalanced before implementing the matching algorithm, well past the recommended threshold of $\pm0.2$. After matching, no covariate exceeds the recommended threshold.}
    \label{fig:loveplot}
\end{figure}

According to the Love plot \citep{prop-match-diagnostics}, presented here in Figure \ref{fig:loveplot}, matching improves the balance among the covariates, except for variables week in season and the signed score difference before and after the run. However, for these three variables, the standardized bias was already near zero, and the change is arguably negligible. Most notably, we observe a marked improvement in the distribution of the propensity scores after matching (see Figure \ref{fig:loveplot} and panel (b) of Figure \ref{fig:covbal-propscore}). Further, the standardized bias for each variable is less than 0.2 in absolute value, suggesting no covariate imbalance in the matched cohort \citep{stuart2010matching}.

To rigorously assess the balance of the matched cohorts, hypothesis testing is used to check for discrepancy in each of the covariates listed in Table \ref{covariate-list} as well as the estimated propensity score. For discrete and continuous variables, bootstrapped Kolmogorov-Smirnov tests were applied, shown to have correct coverage in \cite{abadie2002bootstrap}. Multiple comparison correction is performed to control the false discovery rate at $0.05$ \citep{benjamini1995controlling}. For binary variables, $t$-tests are used, and chi-squared tests are used for categorical variables. Before matching, a discrepancy between the distribution of covariates associated with treatments and controls was identified in all covariates listed in Table \ref{covariate-list} except for week in season (see Section S5 of the \hyperlink{suppA}{Supplement}). A distributional discrepancy was also identified in the estimated propensity score. After matching, there was no evidence of a discrepancy in covariate distributions for any of the listed covariates or the estimated propensity score. Hence, genetic matching appears to yield a matched cohort similar in both the covariates and the propensities.

\subsection{Outcomes} \label{sec-outcome}

The goal of the outcome measure is to quantify the BiT team's response to a run following a timeout (or lack thereof). Naturally, we first considered the change in the score difference from the time of treatment to the end of the post-treatment window. This approach, however, ignores any mid-window scoring, potentially treating two fundamentally different responses the same simply because they share the same start/end points. For example, suppose the BiT team scores five points after the time of treatment followed by five points by the opposing team.  This would correspond to no change in the score difference, and is thus equivalent to a situation when neither team scored during the post-treatment window. 

To address this concern, we subsequently considered the most extreme change in the score difference occurring in the post-treatment window.  Unfortunately, this method suffers from a similar problem as described above. For example, suppose the opposing team continued their run in the post-treatment window, scoring five points before the BiT team answered with a three-point shot. This response is treated the same as if the opposing team scored five unanswered points in the post-treatment window and the BiT team scored zero. Finally, we considered using the change in the BiT team's win probability during the post-treatment window \citep{yam2019}, but such an approach relies on the unknown model used to produce this measure, is time variant over the course of a game, and is heavily dependent on the score difference at the time of the treatment.

\begin{figure}[hbtp!]
    \centering
    \includegraphics[scale=0.6]{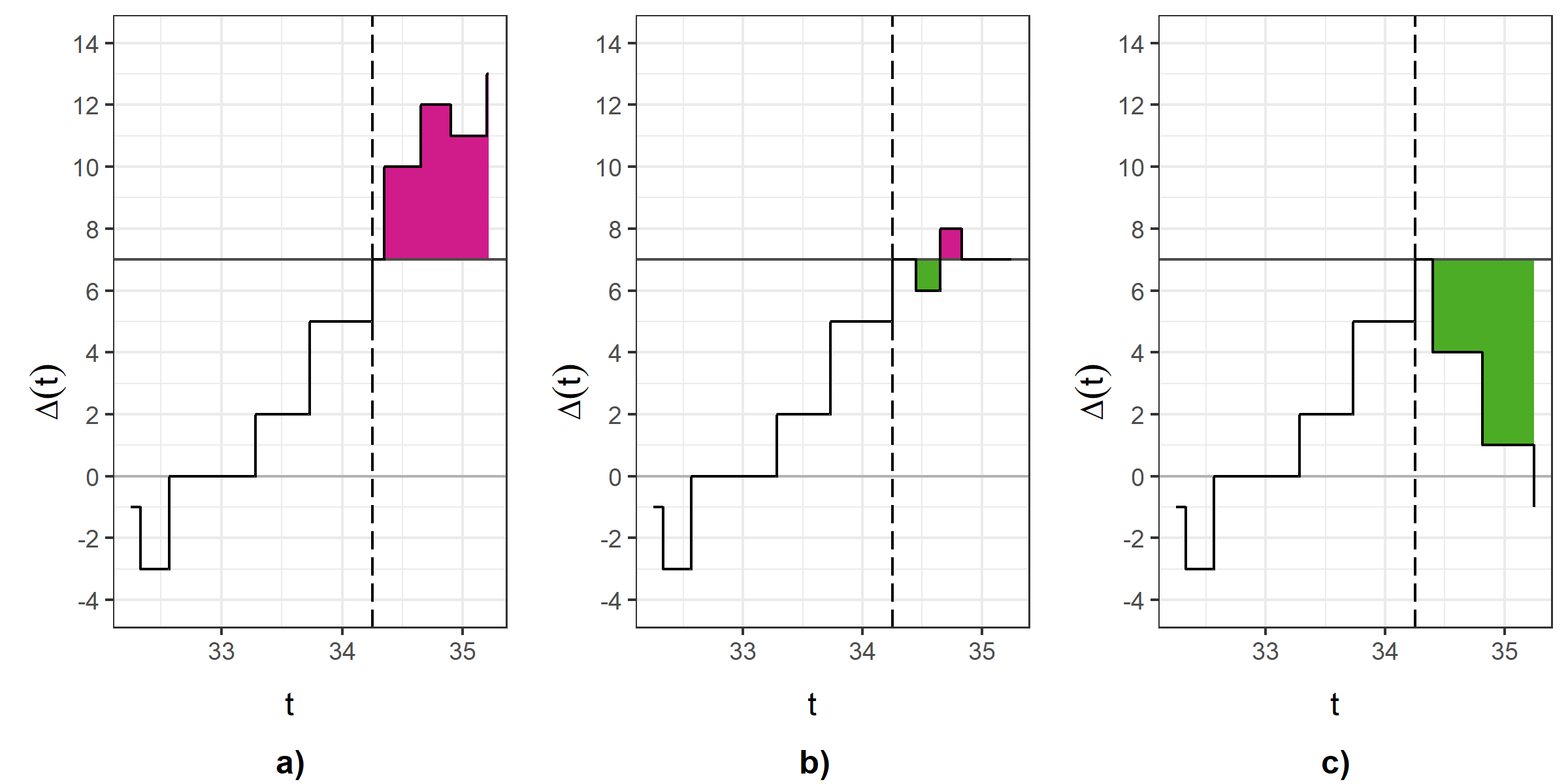}
    \caption{The shaded area represents the integral in Equation \eqref{outcome}. Panel (a) shows the score difference changed at a rate of 3.47 points per minute for the opposing team during the post-treatment window, signifying a negative response to the run following the timeout. Panel (c) shows a score difference that changed at a rate of 4.08 points per minute for the BiT team, a positive response to the run following the timeout. In the panel (b), the opposing team and the BiT team swapped points at a near even rate, neither a negative nor a positive response to the run following a timeout.}
    \label{fig:toy-outcome}
\end{figure}

The clear drawbacks of the initially proposed outcome measures suggested a new method for quantifying a team's response during the post-treatment window was needed.  Our goal is to capture how the score difference holistically changes in the entirety of the post-treatment window without punishing (or rewarding) the BiT team for the score difference at the time of treatment. To this end, we develop the outcome for the $i^{th}$ unit, denoted $y_{i}$, as the integrated, centered-score difference, defined by
\begin{equation} \label{outcome}
    y_{i}= -\sgn{\big(s(t_i)\big)}\int_{t_i}^{t_i+1}{\Big[ \Delta(x) - \Delta(t_i) \Big]}dx,
\end{equation}
where $t_i$ is the time of the treatment for the $i^{th}$ unit, $s(t_i)$ is the signed run point total at time $t_i$, and $\sgn$ is the sign function. Notice that the integrand is the score difference centered by that at the time of treatment. This ensures we do not penalize (or reward) the BiT team for play occurring before the time of treatment. This outcome measures the BiT team's response to the opposing run following an intervention (either a timeout or no timeout), such that positive values indicate evidence of stopping the run, zero indicates a scoreless response or an even exchange in scoring, and negative values indicate evidence of a continued run. Figure \ref{fig:toy-outcome} shows the outcome for three different units on the same score difference curve.  

If the home team is on the run at time $t_i$ then the score difference had a positive trajectory in the pre-treatment window.  In this case, a continuation of this increasing trend in the post-treatment window signifies a negative response to the treatment (see panel (a) of Figure \ref{fig:toy-outcome}).  Conversely, if the score difference started decreasing after the time of treatment, this would indicate a positive response by the BiT team to the treatment (see panel (c) of Figure \ref{fig:toy-outcome}). The negative sign in the outcome definition in Equation \eqref{outcome} ensures the measure aligns with intuition: the first scenario has a negative outcome and the second scenario has a positive outcome.

The sign of the signed run point total is used to create an interpretable outcome, regardless of which team is on the run at time $t_i$.  Contrary to the example just given, if the away team is on the run at time $t_i$, then the score difference is trending negatively in the pre-treatment window and the signed run point total would be negative, $s(t_i)<0$.  The sign function cancels the negative sign in the front of Equation \eqref{outcome}, which is desirable as positive trajectories in the score difference in the post-treatment window then result in a positive outcome. If the outcome is zero, there is either no scoring in the post-treatment window or the BiT team exchanged points with the opposing team at an even rate.

\begin{figure}[hbtp!]
    \centering
    \includegraphics[scale=0.6]{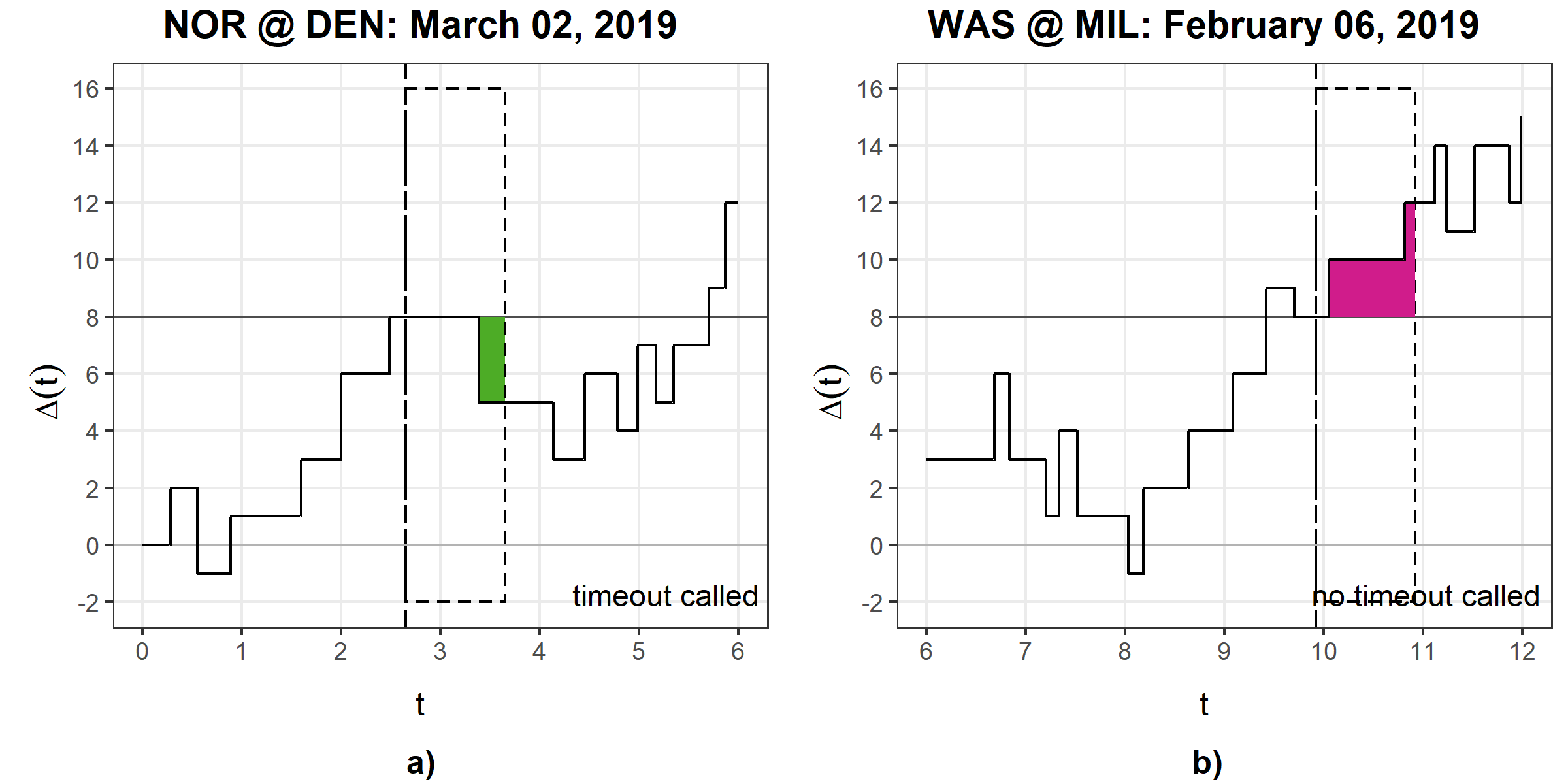}
    \caption{The outcome for the timeout highlighted in Figure \ref{fig:runplot} is shown in panel (a), and the matched control is shown in panel (b).  In panel (a), the New Orleans Pelicans positively responded to the Denver Nuggets run when a timeout was called; however, within the selected match, the Washington Wizards negatively responded to the Milwaukee Bucks' run when no timeout was called. The dashed regions outline the post-treatment window for each unit.}
    \label{fig:outcome}
\end{figure}

The outcome for the example highlighted in Figure \ref{fig:runplot} is 0.80 and is shown again in panel (a) of Figure \ref{fig:outcome}. Similarly, the outcome for its selected match is -1.93 and is shown in the panel (b) of Figure \ref{fig:outcome}. The key desirable property of this outcome is its ability to classify the response to the timeout on a spectrum where negative values indicate failure in stopping the run, zero indicates no change of score or an even exchange of scoring, and positive values indicate success in reversing the run. The larger the magnitude of the outcome, the more extreme the response to the run.

\section{Results}
\label{results}

\begin{figure}[hbtp!]
    \centering
    \includegraphics[scale=0.6]{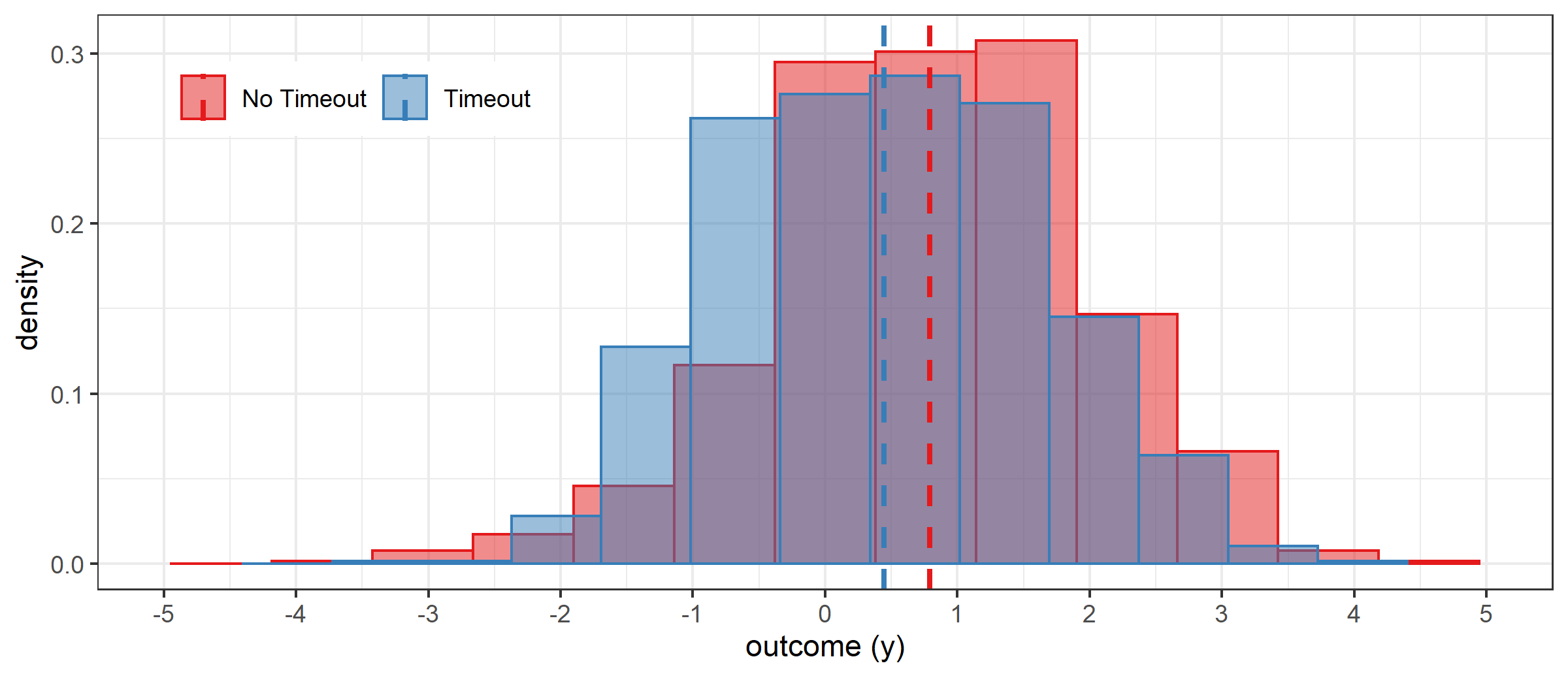}
    \caption{The distribution of the outcome for each intervention group within the matched cohort. The estimated average treatment effect of \att{} is given by the mean of the treated group (blue vertical line) minus the mean of the control group (red vertical line).}
    \label{fig:p-dens-outcome}
\end{figure}

With a balanced, matched cohort of treated and control units, we turn to estimating the causal effect of a timeout. Histograms for the outcomes in the treatment and control groups are given in Figure \ref{fig:p-dens-outcome}. First, note that a large portion of the outcome distribution for both the treated and control cohorts is greater than zero, indicating that the BiT team often has some level of a comeback, regardless of whether there is stoppage in play. We interpret this as evidence that momentum shifts are common, and thus should be expected. 

The estimate of the average treatment effect on the treated is visualized by the difference between the mean of the treated group and the mean of the control group of the matched cohort (shown by the dashed lines in Figure \ref{fig:p-dens-outcome}). The estimated average treatment effect on the treated is \att{} with an Abadie-Imbens standard error \citep{abadie2004implementing} of 0.07 and an associated $p$-value of $p$ \attPvalue{}. Hence, on average, it appears slightly disadvantageous to call a timeout in the presence of an opposing run. While these results may seem counterintuitive, the negative estimate aligns with the unmatched, naïve estimate of -0.08. This measured effect, however, is insignificant until employing formal causal methodology to account for the inherent self-selection bias associated with observational studies.

Due to the non-deterministic nature of the matching algorithm, we re-ran the matching step in our analysis to study the variability in the estimated ATT due to different matched cohorts. Twenty realizations of the estimated ATT are presented in Section S3 of the \hyperlink{suppA}{Supplement}, indicating that results were consistent across different matched cohorts.

Acknowledging that the effectiveness of a timeout in stopping an opposing run may vary across teams due to, say, the maturity of the players or the ability of the coach to strategize a comeback, we estimated the ATT for each franchise individually. The average treatment effect on the treated for franchise $f$, $ATT_{f}$, is defined 
\begin{equation} 
\label{eqn:attf}
ATT_f = 
\Exp\left[ Y_{i}(1) - Y_{i}(0) \mid T_{i} = 1, \mathcal{B}_i = 1 \right],  \\
\end{equation}
where $\mathcal{B}_i=\mathbb{I}_{\{f \text{ is the BiT team for the } i^{th} \text{ unit}\}}$. The matching procedure does not guarantee covariate balance within franchise. Therefore, we employ a hypothesis testing approach to assess whether or not the covariates are sufficiently balanced when conditioning on the BiT team's identity (see Section S5 of the \hyperlink{suppA}{Supplement}). After controlling the false discovery rate at 0.05 to account for multiple comparisons \citep{benjamini1995controlling}, there are no significant findings, indicating it is reasonable to assume the treatment is ignorable when conditioning on the BiT team's identity.

\begin{figure}[htbp!]
    \centering
    \includegraphics[scale=0.7]{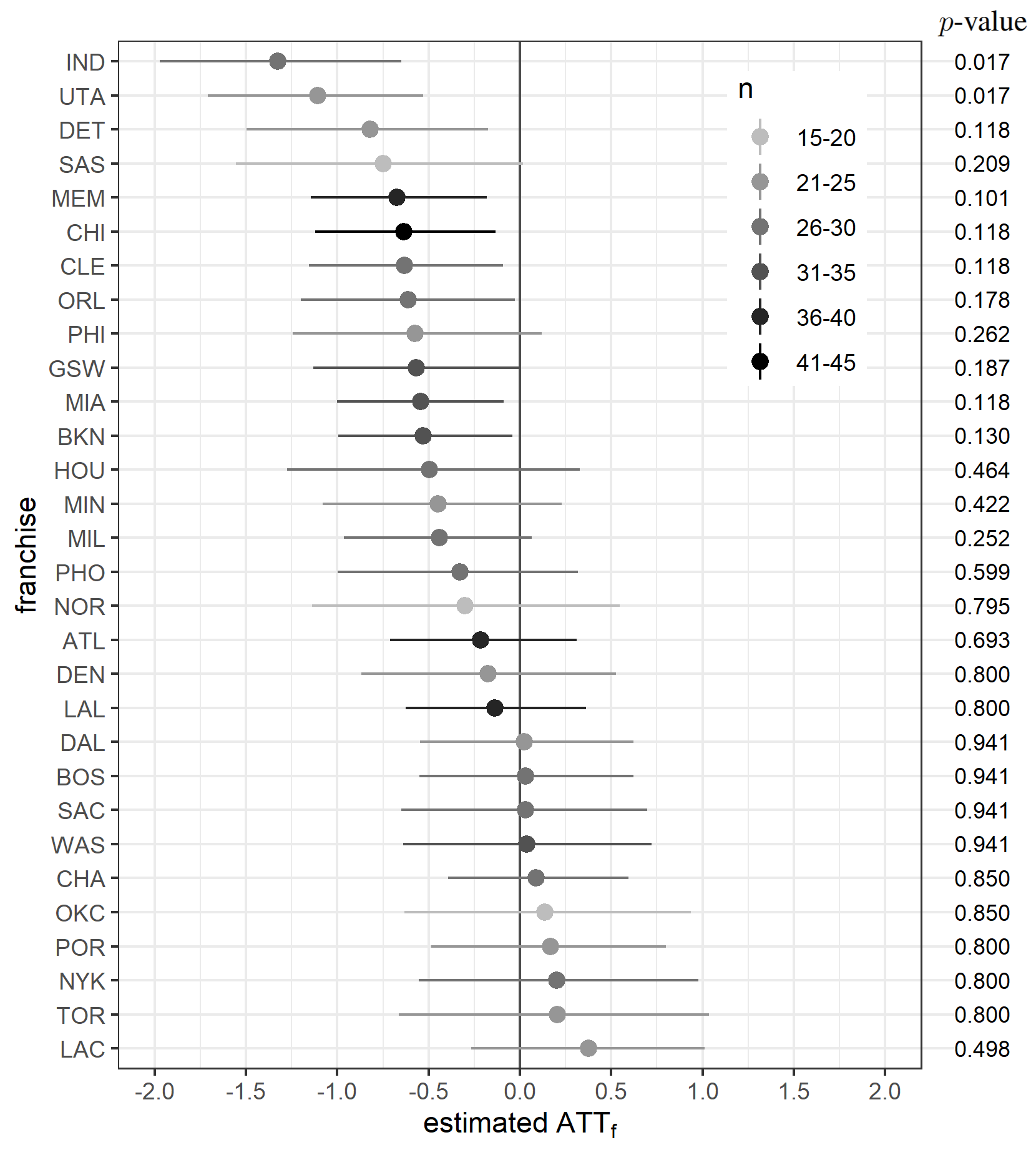}
    \caption{Estimated average treatment effect for the treated for each franchise with a 95\% confidence interval based on a non-parametric bootstrap. Ignoring the multiplicity problem, there are ten significant results at a significance level of 0.05. Computing the unadjusted $p$-values with paired permutation tests and controlling the false discovery rate at 0.05, the adjusted $p$-values are provided on the right margin. After accounting for multiple testing, the Indiana Pacers and the Utah Jazz have significant, negative average treatment effects on the treated. For these franchises, the opposing team scores at a significantly faster rate concluding a timeout than when a timeout is not called during an opposing run.}
    \label{fig:attf}
\end{figure}

To estimate $ATT_f$ for each franchise, the matched treated and control units were partitioned based on the BiT franchise of the treated unit and the average within-matched-set mean differences in outcome was computed. Some franchises obviously were associated with more treated units than others (see Section S4 of the \hyperlink{suppA}{Supplement}), so bootstrapped samples were created to quantify the variability of the causal estimator in Equation \eqref{eqn:attf}, as in \cite{yam2019}. The results are given in Figure \ref{fig:attf}.

Immediately, we notice that twenty of the thirty franchises exhibit negative point estimates for their franchise average treatment effect on the treated. For these franchises, the negative estimated effect suggests, on average, the opposing team scores at a faster rate when a timeout is called than when it is not called during an opposing run. To assess statistical significance of these estimates and address the multiplicity problem, paired permutation tests are conducted for each of the franchises, and the two-sided $p$-values are recorded. After computing the unadjusted $p$-values, we employ the strategy suggested in \citet{benjamini1995controlling} to control the false discovery rate at 0.05. 

After accounting for multiple comparisons, there are two significant, negative franchise treatment effects on the treated: those corresponding to the Indiana Pacers and the Utah Jazz. Relatively small sample sizes may have contributed to the lack of significant findings after accounting for multiple testing. None of these positive effects withstand statistical significance. 

\section{Sensitivity Analysis}
\label{sec:sensitivity}

 A sensitivity analysis is employed to assess the robustness of the results (1) with regard to the specification of a run, and (2) in the presence of unobserved confounding. In this work, a run is characterized not only by the magnitude of the change in the score difference (run point total) but also the time taken by the opposing team to attain said change (run duration). While few would argue the importance of these two attributes in defining a run, there is no universal agreement on what values constitute a run. We define a play to be a run if it features a nine-point change in the score difference attained in the prior two minutes. Considering the average NBA possession is conservatively 15 seconds \citep{beuoy}, a two-minute window allows for approximately eight possessions, or four possessions per team. The current definition was motivated by the simple scenario: four defensive stops and at least three three-point shots made yields a drastic change in the game in a short amount of time. We acknowledge that this definition is arbitrary and should be explored more carefully. Therefore, we examine the sensitivity of our results relative to different combinations of run point total thresholds (7, 8, 9, and 10 points) and limits to the run duration (1.5, 2, 2.5, 3). For each combination, we replicate the analysis and report the findings in Table \ref{tbl:stparams-results} of Appendix \ref{app:sens}. As can be seen, the estimated effect is negative and significant for each combination, indicating robustness to alternative run definitions.

\begin{figure}[htb!]
    \centering
    \includegraphics[scale=0.73]{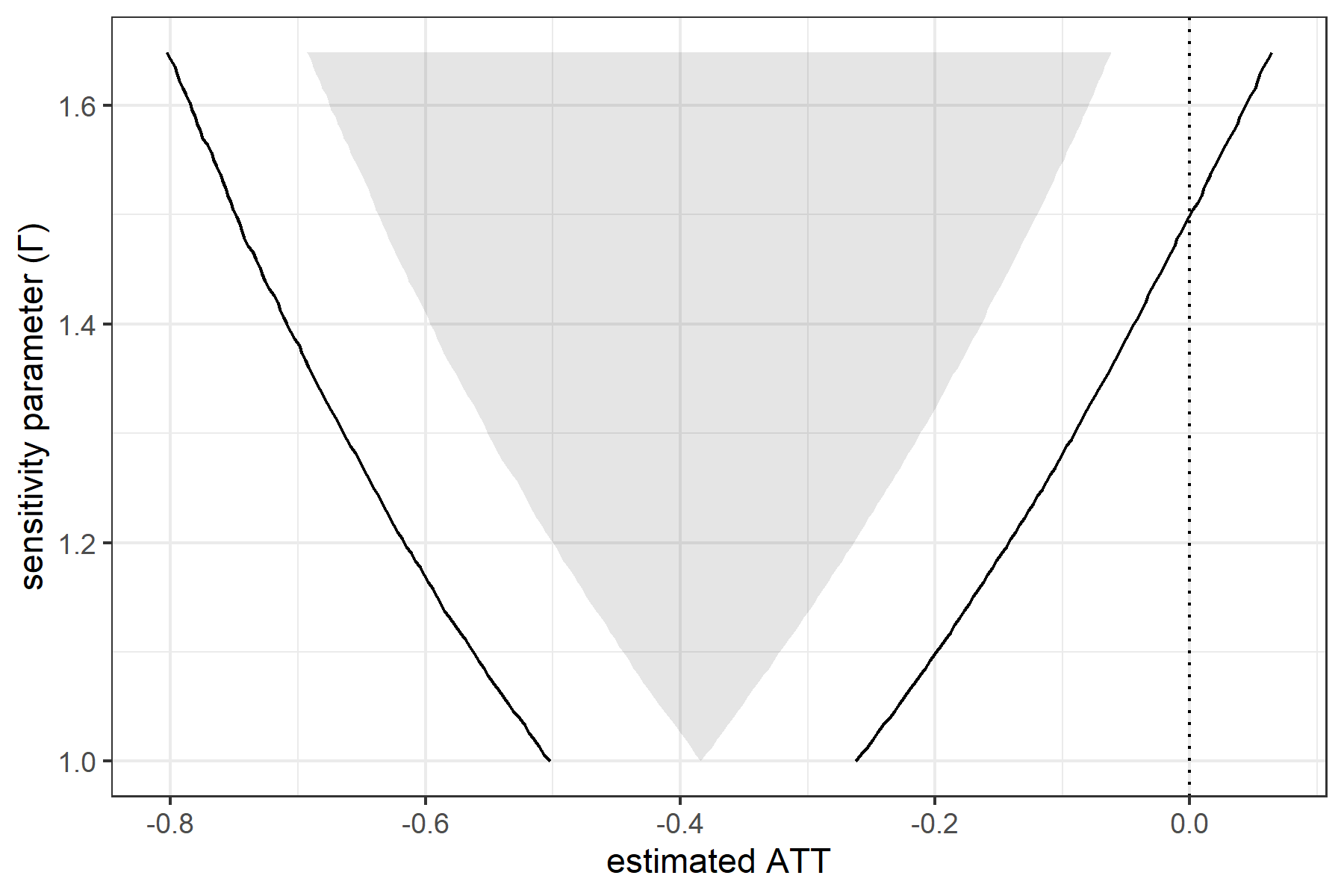}
    \caption{Provided a bias of $\Gamma$ in the treatment assignment, the shaded region illustrates the interval of point estimates possible, and the solid black lines illustrate the (conservative) 95\% confidence interval. The confidence interval includes 0, indicating no effect, at $\Gamma\approx 1.50$, the magnitude of bias from nonrandom assignment necessary to alter the conclusions herein. While these results indicate the study may be sensitive to unobserved confounding, the degree to which is likely overstated \citep{rosenbaum2015two}.}
    \label{fig:rossen}
\end{figure}

Aside from the run specification, it is important to consider the robustness of the perceived effects to unmeasured or unobserved confounders. While matching methods can adjust for observed confounding (assessed through covariate balance before and after matching), the impact of unobserved confounding on the estimated effect must be explored through a sensitivity analysis \citep{rosenbaum2007sensitivity}. Following the recommendations of \cite{rosenbaum2013impact}, we let $\Gamma$ represent the magnitude of the bias from nonrandom treatment assignment and estimate the ATT and confidence interval for the ATT as in Equation \ref{eq:matchingATT} at various $\Gamma$. The results are illustrated in Figure \ref{fig:rossen}. When $\Gamma\approx 1.50$, the confidence interval for the treatment effect includes 0, indicating a materialistic change in the inferential conclusions of this study. While these results imply sensitivity to unobserved confounding, it is important to note that the 95\% confidence interval is known to be conservative \citep{rosenbaum2015two}. As stated by \citet{liu2013introduction}, this method assumes a situation in which ``the unobserved confounder perfectly predicts the outcome of interest," an unrealistic assumption in practice. As a result, sensitivity to unobserved confounders is likely overstated.

\section{Discussion}
\label{discussion}

While the idea of a ``run'' is commonly used within the context of basketball, there is no formal mathematical definition. Part of the novelty of this work is formalizing the colloquial understanding of a run as it pertains to professional basketball, while developing an interpretable outcome which captures the relative performance of each team in a game where the score changes frequently. After proposing a framework with which to study runs, we employed causal methods to estimate the potential gain (or loss) attributed to a timeout during a team's run. We find that, on average, calling a timeout worsened the non-run team's short-term performance compared to if no timeout was taken during an opposing run. In particular, the Indiana Pacers and the Utah Jazz short-term performance significantly declines from a timeout compared to if no timeout was taken, on average. No teams, on average, exhibit a significant gain in their short-term performance from a timeout compared to if no timeout was taken.

One important variable which is not considered within this analysis is substitutions that occur when a timeout is called. These are reasonably assumed to have a positive impact on the outcome but cannot be included within the matching procedure since these substitutions occur after or simultaneously with the treatment. That said, more substitutions actually exist within the group of treated units, when a timeout is called, than in the group of control units, when a timeout is not called. This suggests the effect of a timeout may actually be less than that estimated in this analysis. 

There are choices made in defining units which could be addressed by generalizing the causal methods used. For example, plays occurring in the last minute of a period are excluded from this analysis. Since the outcome requires a succeeding one minute of game time after intervention, any play occurring in the last minute of a period is akin to a censored observation and is thus omitted. Currently, for a given play at time $t$, if there are timeouts in the pre-treatment window (two minute interval of time prior to $t$), then that play is excluded from the analysis on the basis of multiple variations. The number of timeouts and the time between timeouts could be used to create a more general, non-dichotomous treatment regime and is an area of future work. 

Causal inference is becoming a popular tool for sports analysts due to the observational nature of sporting events. However, despite the rise in popularity, too often little emphasis is placed on closely examining whether the stable unit-treatment-value assumption (SUTVA) is reasonable. In this work, we prioritize this discussion by not only defining units but describing how their definition was motivated by adherence to SUTVA. Aside from the honest exposition surrounding SUTVA, we also take great care in defining an interpretable outcome which measures relative short-term performance well. This outcome is flexible and well suited for applications where the quantity to measure is susceptible to short term fluctuations, but the interest is in the average change. Other phenomena that might benefit from the modeling tools introduced here include stock prices, approval ratings, and of course, score differences in professional sports.

\section*{Acknowledgements}
We would like to thank Damian Wandler for initially posing an early variant of this problem to one of the authors. In addition, we thank the organizers and attendees of the 2019 Carnegie Mellon Sports Analytics Conference for the insightful questions, comments, and suggestions, and the organizers of the 2019 UConn Sports Analytics Symposium for funding which made presenting this work at its early stages possible.  Finally, we thank the anonymous Associate Editor and reviewer for their insightful comments that led to substantial improvements in the work.

\bibliographystyle{apalike}
\bibliography{references}

\newpage

\appendixtitleoff
\begin{appendices}
\bigskip
\noindent{\Large\textbf{Appendix} \par}

\section{Characterizing SUTVA}
\label{app}

\begin{figure}[hbtp!]
    \centering
    \includegraphics[page=1,width=1\textwidth]{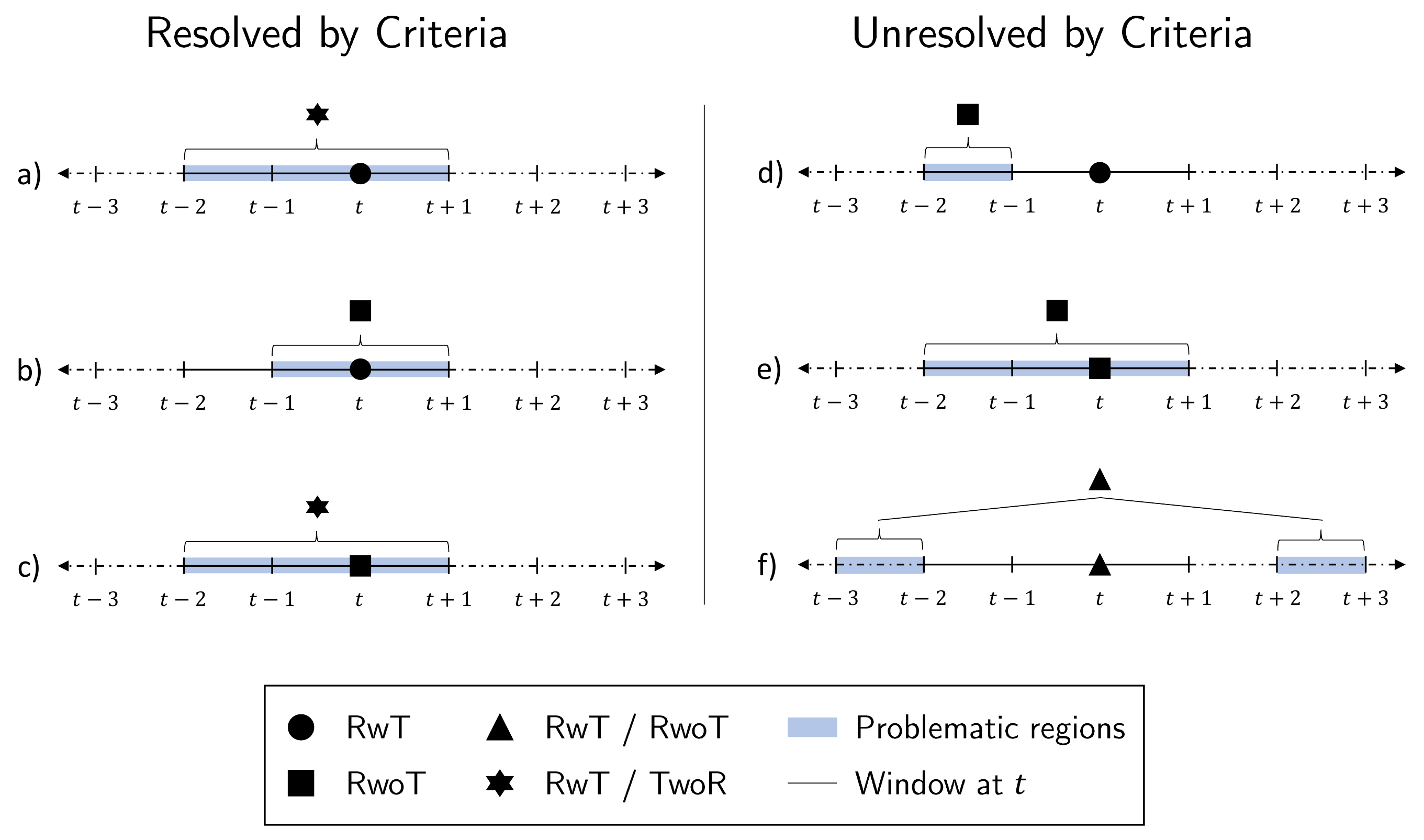}
    \caption{Possible violations of SUTVA for a fixed play at time $t$. The criteria resolves issues with interference and hidden variation pertaining to runs with timeouts (RwTs), being the treatment units; however, SUTVA also requires non-interfering runs without timeouts (RwoTs), being the control units. This assumption is relaxed to adequately match of time-dependent covariates such as run duration and time left in the game and to maintain a reasonable sample size for the analysis.}
    \label{fig:interference_graph}
\end{figure}

To address violations of the SUTVA, criteria (2) and (3) in the unit definition in Section \ref{methods-treatments-and-controls} were introduced. These criteria handle issues with hidden variation and interference attributed to plays involving a timeout occurring in the pre-treatment or post-treatment window. While mitigating serious violations to the SUTVA, these criteria allow a reasonable control pool for matching on time-dependent covariates. All violations to the SUTVA are provided in Figure \ref{fig:interference_graph}, grouped by those cases which are resolved and unresolved by the current criteria. 

By removing instances which feature a timeout in the pre-treatment or post-treatment window, we largely address concerns with overlapping windows among those units in the treated group (Figures \ref{fig:interference_graph}\textcolor{blue}{.a} , \ref{fig:interference_graph}\textcolor{blue}{.b}, and \ref{fig:interference_graph}\textcolor{blue}{.c}) and curtail issues with hidden variations (Figure \ref{fig:interference_graph}\textcolor{blue}{.a}). The only complication remaining are overlapping instances where the intersection between the windows is relatively small (Figures \ref{fig:interference_graph}\textcolor{blue}{.d} and \ref{fig:interference_graph}\textcolor{blue}{.f}).

Furthermore, units in the control group still experience overlapping windows (Figure \ref{fig:interference_graph}\textcolor{blue}{.e}). To resolve the issue with overlapping control units, controls would need to be removed until the remaining set of controls maintain mutually disjoint windows. Clearly, this set is not unique and defining the optimal set of controls to remove is unclear. In contrast, maintaining overlapping controls allow for more precision when matching on time-dependent covariates. For example, seconds after an instant of a RwoT is likely also a RwoT; however, these two RwoTs differ in their run duration, calculated by Equation \ref{eqn:run-duration}. If we were to remove one of these instances to ensure the pool of controls do not overlap, then matching on time-dependent covariates such as run duration and time left in the game may suffer. As a result, we recognize the slight violation of SUTVA so not to introduce bias in the matching procedure.

After matching, we investigate the degree of overlapping windows in the matched cohort of controls. Since the matching procedure allows a control to serve as the matched counterfactual for more than one treatment, then the controls are not necessarily unique. Of the \nTreatments{} matched controls, there are 617 unique controls. Of the 617 unique controls included in the matched cohort, the maximal number of disjoint windows we can construct from these 617 unique controls is 561. So, of the 617 unique controls, at most 561 of them have non-overlapping windows. While we allowed for overlapping windows in the group of controls, after employing the matching algorithm, the number of overlapping windows within the cohort of matched controls is minor.

\counterwithin{table}{section}

\section{Sensitivity to Run Definition}
\label{app:sens}

The analysis herein is replicated for various definitions of a run obtained by considering four different run point totals and four different pre-treatment window lengths (see Section \ref{sec:sensitivity} and Table \ref{tbl:stparams-results}). Regardless of specific choice in run definition, the estimated causal effect of a timeout is negative and significant, aligning with the presented results. This indicates that the results are robust to the characterization of a run.

\begin{table}[ht!]
\centering
\caption{Results of the analysis for various definitions of a run. In particular, the inference drawn is relatively the same, illustrating robustness of the results to the characterization of a run.}
\begin{tabular}[t]{lllrrrrrrrl}
\toprule
\multicolumn{2}{c}{} & \multicolumn{3}{c}{N} & \multicolumn{3}{c}{Results} \\
 \cmidrule(l{3pt}r{3pt}){3-5} \cmidrule(l{3pt}r{3pt}){6-8}
Points & Window & Units & RwT & RwoT & $\widehat{ATT}$ & $SE$ & $p$ & \\
\midrule
7 & 1.5 & 20,603 & 2,626 & 17,977 &  -0.23 & 0.04 & $<0.001$ & \\
7 & 2.0 & 34,005 & 3,694 & 30,311 & -0.25 & 0.03 & $<0.001$ & * \\
7 & 2.5 & 41,520 & 4,285 & 37,235 & -0.28 & 0.03 & $<0.001$ & * \\
7 & 3.0 & 43,821 & 4,472 & 39,349 & -0.26 & 0.03 & $<0.001$ & * \\
\addlinespace
8 & 1.5 & 6,816 & 1,090 & 5,726 &  -0.38 & 0.06 & $<0.001$ & \\
8 & 2.0 & 13,677 & 1,911 & 11,766 &  -0.31 & 0.05 & $<0.001$ & \\
8 & 2.5 & 18,868 & 2,470 & 16,398 &  -0.34 & 0.04 & $<0.001$ & \\
8 & 3.0 & 21,618 & 2,745 & 18,873 & -0.26 & 0.04 & $<0.001$ & * \\
\addlinespace
9 & 1.5 & 1,600 & 325 & 1,275 &  -0.37 & 0.14 & 0.010 & \\
9 & 2.0 & 4,684 & 834 & 3,850 &  -0.35 & 0.07 & $<0.001$ & \\
9 & 2.5 & 7,547 & 1,204 & 6,343 &  -0.31 & 0.06 & $<0.001$ & \\
9 & 3.0 & 9,572 & 1,459 & 8,113 &  -0.28 & 0.05 & $<0.001$ & \\
\addlinespace
10 & 1.5 & 370 & 103 & 267 &  -0.51 & 0.20 & 0.009 & ** \\
10 & 2.0 & 1,446 & 297 & 1,149 &  -0.50 & 0.16 & 0.001 & \\
10 & 2.5 & 2,923 & 532 & 2,391 &  -0.43 & 0.13 & 0.001 & \\
10 & 3.0 & 4,139 & 710 & 3,429 &  -0.44 & 0.09 & $<0.001$ & \\
\bottomrule
\end{tabular}
\parbox[t]{\textwidth}{\footnotesize
  $^*$~match tolerances were relaxed after run time exceeded seven days (see Appendix \ref{app:gen_match}) \newline
  $^{**}$~propensity score model failed to converge
}
\label{tbl:stparams-results}
\end{table}

\section{Genetic Matching Algorithm}
\label{app:gen_match}

The genetic matching procedure used to construct a matched cohort has several arguments that dictate the speed of the process. A few of these arguments include the the population size, wait generation, maximum generation, and the distance tolerance \citep{genmatchingpackage}. Since the theory proving genetic matching yields reasonable solutions is asymptotic in the population size, generational change yields improvement in optimization, and the distance tolerance defines the closeness of a proposed match, the population size and generations are set reasonably high while the tolerance is set reasonably low. In particular, the arguments used are a population size of 8,000, wait generation of 4, max generation of 100, and tolerance of 0.00001. At these values, the optimization routine yields a solution before the generational limits are met. However, for definitions of a run yielding many observations (marked by * in Table \ref{tbl:stparams-results}), the optimization routine was manually stopped after running for one week. At this point, we relaxed the arguments until the routine reached a run time less than one week. The following arguments are used for such run definitions: population size of 1,000, wait generation of 2, max generation of 15, and tolerance of 0.1. At these relaxed constraints, the optimization routine yielded a solution after nearly three days. On the other hand, the run definitions yielding relatively few observations (marked by ** in Table \ref{tbl:stparams-results}) induced complications in estimation of the propensity score. When fitting the generalized additive model discussed in Section \ref{sec-prop-score} to this data set, the procedure failed to converge. We recommend caution in interpreting results demarcated by asterisks.
\end{appendices}

\newpage

\appendixtitleoff
\begin{appendices}
\renewcommand{\thesection}{S.\arabic{section}}
\renewcommand{\thefigure}{S.\arabic{figure}}
\renewcommand{\thetable}{S.\arabic{table}}
\setcounter{section}{0}
\setcounter{figure}{0}
\setcounter{table}{0}
\bigskip
\noindent{\large\textbf{\hypertarget{suppA}{Supplement}} \par}
  
\section{Data Preparation and Manipulation}

After gathering play-by-play data for the 2017-18 and 2018-19 seasons, data preparation is needed before defining and identifying runs. We start by collapsing multi-row plays (\emph{e.g.}, a foul and free throws) into one row. This is achieved by recording whether a timeout was called when the clock is stopped and retaining the last recorded scoring event. If there were no scoring events, the last event is retained. In exploring the data, we find that some rows are simply empty and others correspond to a game which is clearly misreported, taking on an infeasible score in the time reported. These rows are removed from the data set. Since overtime is not included in the analysis, rows timestamped after the fourth period are also removed. Furthermore, the data set is limited to instances in time when the score keeper noted some change in the game; however, time is continuous and ought to be treated as such. To this end, we expand the play-by-play data set to contain rows every five seconds. This approach is not suspect since it is safe to assume that the score remains constant if no change in the score is reported. This allows us to more aptly capture runs without timeouts. Lastly, pseudo-plays (dubbed plotting plays) are rows added to the data set before and after any play dictating a change of score in the data set. These plays cannot be considered a unit (removed from consideration by criteria 2 and 3), and are only used to plot the score differential over time, making it a step-wise function to reflect the instantaneous change in score. If a play at time $t$ marks a change in the score difference, then a plotting play is added at $t-\rho$ with the score difference prior to $t$ and another is added at $t+\rho$ with the score difference at $t$ where $\rho$ is arbitrarily close to zero.

After preparing the data set for analysis, every play is evaluated based on Equation (2) of the manuscript. Any play maintaining a signed run point total of \text{NA} is removed. As such, only run plays remain, and evaluating whether a timeout was called during that run is now possible. To quail concerns with SUTVA violations, we introduce criteria (2) through (4). We start by removing any run play containing the beginning or end of the period in pre-treatment or post-treatment window (criterion 4) before removing any run play containing a timeout in the pre-treatment or post-treatment window (criteria 2 and 3).

Upon addressing concerns with SUTVA violations, we turn to potential issues with positivity by removing any run play with a moneyline greater than 2400 in absolute value. After their removal, we are left with 4,684 run plays which are deemed units for the analysis. Of these run plays, 834 are runs with a timeout (RwTs) and 3,850 are runs without a timeout (RwoTs). More details regarding how many observations are removed with each step can be found in Table \ref{tab:obscount}.

{ 
\renewcommand{\arraystretch}{1.4}
\begin{table}[htbp!]
\centering
\caption{Number of observations remaining after each step of the data analysis: including data preparation, criteria for defining units, and positivity concerns. In total, there are 4,684 units, 834 of which are runs with timeouts and 3,850 of which are runs without timeouts.}
\label{tab:obscount}
\begin{tabular}{p{6cm}rR{2.28cm}R{2.28cm}}
 & Observations & Runs with a Timeout & Runs without a Timeout \\
\textbf{Gathering Data} &  &  &  \\ \midrule
\rowcolor[HTML]{EFEFEF} 
Procure data with nbastatR package in R. & 1,144,461 &  &  \\
\addlinespace[4mm]
\textbf{Data Preparation} &  &  &  \\ \midrule
\rowcolor[HTML]{EFEFEF} 
Collapse multi-row plays into one row. & 778,828 &  &  \\
Remove misreported games, overtime plays, and empty rows. & 617,187 &  &  \\
\rowcolor[HTML]{EFEFEF} 
Discretize time to five second intervals. & 2,036,030 &  &  \\
Create plotting plays. & 2,813,946 &  &  \\
\addlinespace[4mm]
\textbf{Invoke Criteria} &  &  &  \\ \midrule
\rowcolor[HTML]{EFEFEF} 
Play must be a run. & 31,081 & 1,149 & 29,932 \\
Windows must be uncensored. & 27,340 & 1,101 & 26,239 \\
\rowcolor[HTML]{EFEFEF} 
Windows must exclude a timeout. & 4,730 & 838 & 3,892 \\
\addlinespace[4mm]
\textbf{Address Positivity} &  &  &  \\ \midrule
\rowcolor[HTML]{EFEFEF} 
Consider moneyline less than 2,400 in absolute value. & 4,684 & 834 & 3,850
\end{tabular}
\end{table}
}

\section{Propensity Score Model}

After compiling the pre-treatment covariates listed in Table 1 of the manuscript, we start by fitting a generalized additive model using all variables but those corresponding to team identity (both the BiT team and the opposing team). Initially, we decided to omit team as a matching covariate since no coach employs a universally consistent timeout strategy (see Treated and Control Franchise Frequency). While some coaches openly claim to strictly adhere to a particular timeout dogma, we've demonstrated this to be false. Our belief is that the act of calling a timeout is predominately driven by the covariates with which we originally constructed the propensity score and conducted the matches. We failed to include team as a matching dimension, feeling an exact match might attenuate the importance of some of the other matching variables. After further investigation, we find that ignoring team identity yields a matched cohort which is significantly unbalanced in terms of the BiT team ($p<0.001$) and marginally unbalanced in terms of the opposing team ($p=0.06$). We decide to investigate the inclusion of these variables in the propensity score model. To this end, we construct three nested propensity score models: one including every covariate in Table 1 of the manuscript, one excluding opposing team, and one excluding both the BiT and opposing team. We conduct two Chi-squared goodness-of-fit tests \citep{young2011generalized, scheipl2008size, wood2013p} for nested GAMs to assess the inclusion of BiT team and opposing team, respectively. We find strong evidence that both BiT team ($p<0.001$) and opposing team ($p<0.001$) should be included in the model.

\setcounter{table}{1}

\begin{table}
\caption{True negative and positive rates (TNR/TPR) and negative and positive predicted values (NPV/PPV) estimated using 70/30\% cross-validation and 1,000 Monte Carlo splits. The full model which yields larger estimated propensities for many units yields a marked improvement in the true positive rate with a negligble decline in the true negative rate. Furthermore, the full model yields marked improvements in both the negative and positive predicted values. This implies a betterment in the proportion of correctly classified treatments and the proportion of classified treatments (controls) which are actually treatments (controls).}
\centering
\label{tab:propscore-rates}
\begin{tabular}[htbp!]{lrrrr}
\toprule
 & TNR & TPR & NPV & PPV\\
\midrule
Restricted Model & 0.987 & 0.089 & 0.834 & 0.599\\
Full Model & 0.974 & 0.185 & 0.847 & 0.612\\
\bottomrule
\end{tabular}
\end{table}

\begin{figure}[hbtp!]
    \centering
    \includegraphics[scale=0.8]{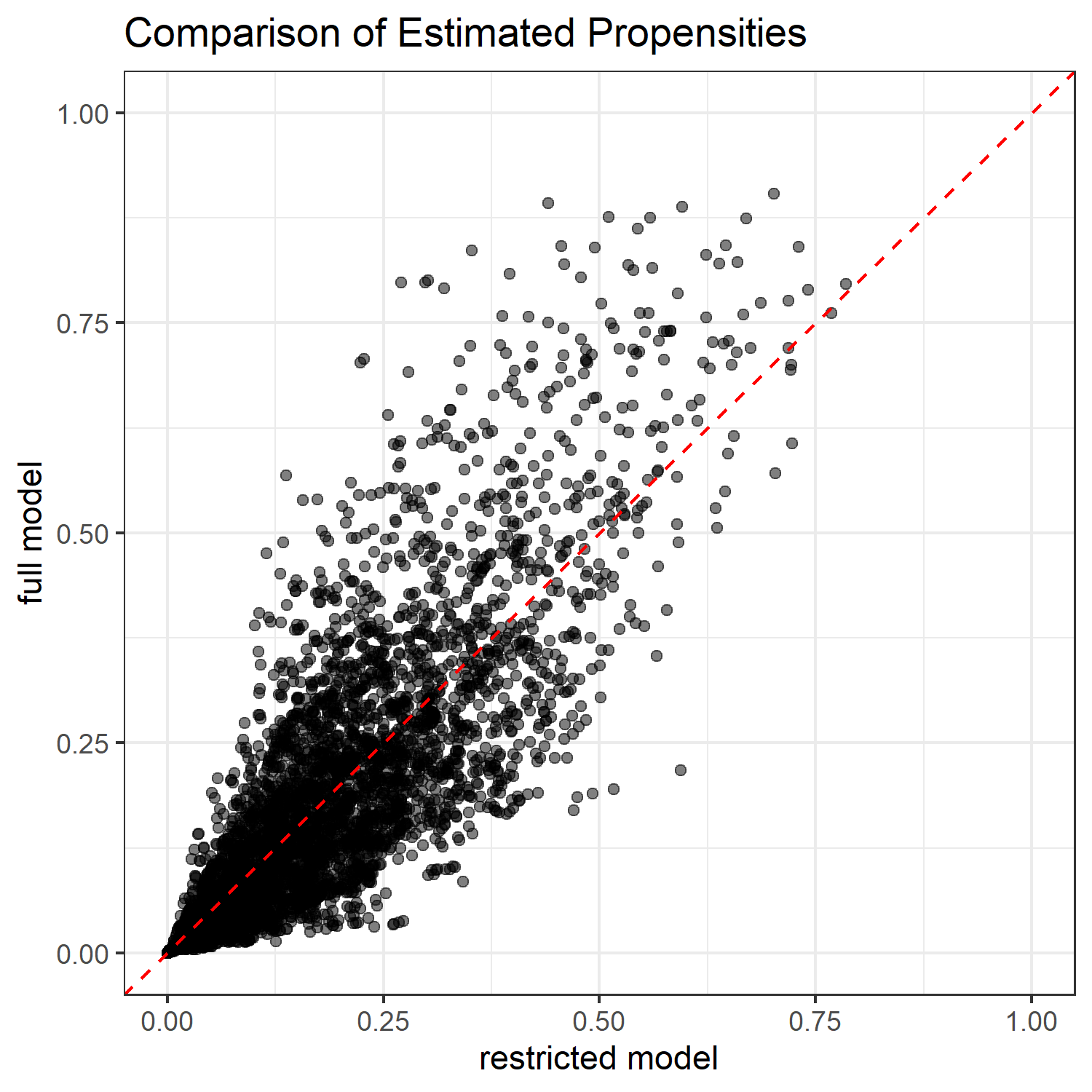}
    \caption{Estimated propensities for the full model (including BiT and opposing teams) and the restricted model (excluding BiT and opposing teams). After including these covariates, more plays are given a larger propensity score, noted by the larger number of observations above the 45 degree line.}
    \label{fig:propscore-comparison}
\end{figure}

Comparing the estimated propensities yielded from the full model (including BiT and opposing teams) and the restricted model (excluding BiT and opposing teams), we notice an increase in the density of larger propensities (see Figure \ref{fig:propscore-comparison}). In general, plays are assigned a higher propensity, indicating more confidence in a called timeout. To investigate the effects of these changes, we employ cross-validation with 1,000 Monte Carlo replicates to estimate the increase (or decrease) in the true positive (negative) rate and the positive (negative) predicted value. For a given replicate, the units are partitioned into a training and testing set according to a uniformly at random 70/30\% split. The restricted and full models are fit using the training set, predictions are made using the testing set, and the metrics are recorded. An average is taken across the 1,000 replicates to estimate the metrics; the results are provided in Table \ref{tab:propscore-rates}. Inclusion of BiT and opposing teams leads to a marked improvement in the true positive rate as well as the negative and positive predicted values with a negligible decline in the true negative rate. Overall, inclusion of these covariates improves the proportion of correctly classified runs with a timeout and the proportion of classified runs with a timeout (runs without a timeout) which are actually runs with a timeout (runs without a timeout).

\section{Matching Variability}

\begin{figure}[hbtp!]
    \centering
    \includegraphics[scale=0.8]{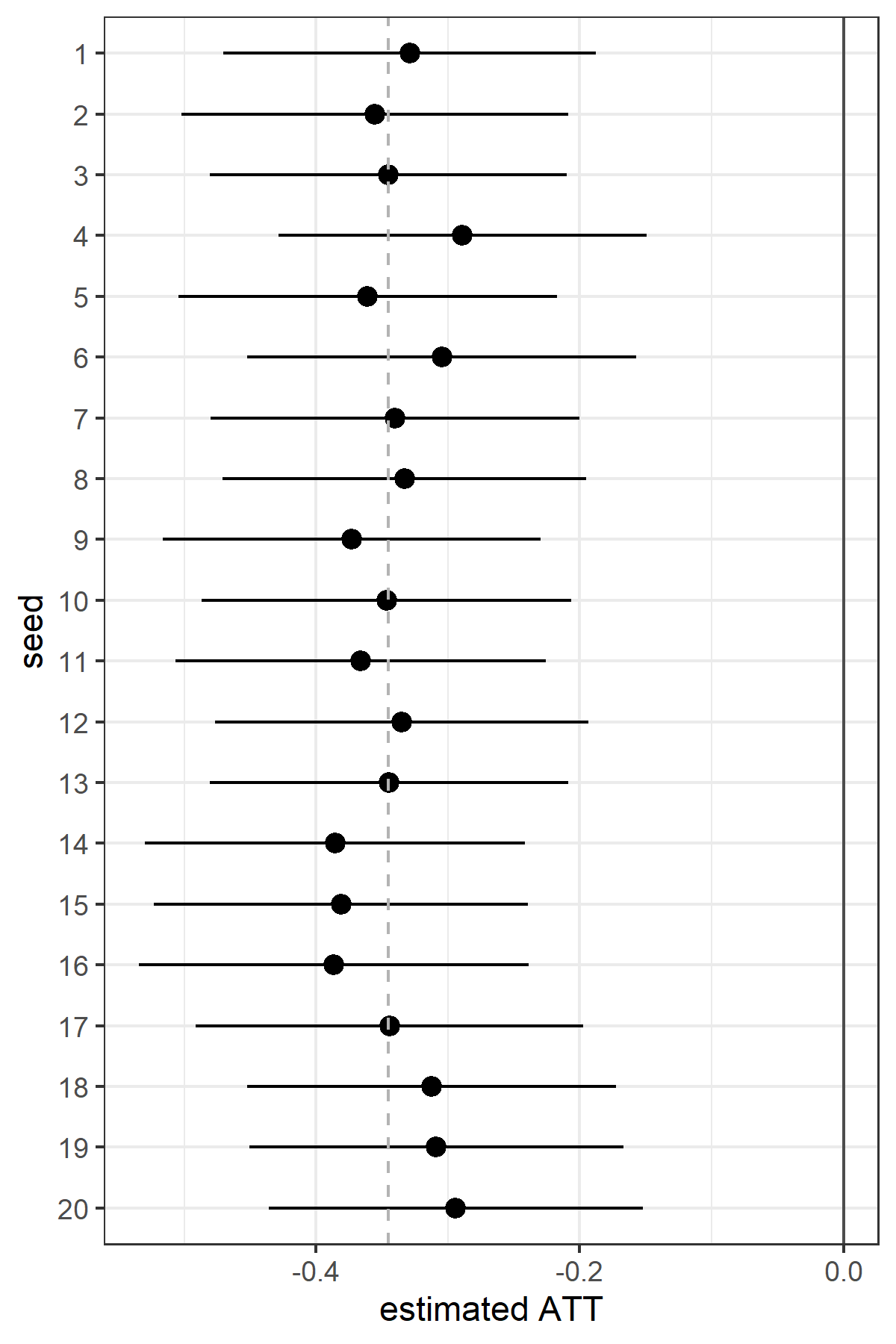}
    \caption{Estimated $ATT$ and 95\% confidence interval for 20 runs of the matching algorithm. The dashed, vertical line represents the estimate used in the paper. Each realization of the matching algorithm produced consistent results.}
    \label{fig:matching-forest}
\end{figure}

Since the genetic matching algorithm provided in the \texttt{R} package \texttt{Matching} is non-deterministic, the matching algorithm was initialized at 20 different random seeds and the $ATT$ was estimated for each of the resulting matched cohorts to explore the variability of the estimator. Figure \ref{fig:matching-forest} shows the $ATT$ estimates and the corresponding 95\% confidence intervals for the twenty matched cohorts. The random seed associated with the estimated $ATT$ closest to the average of the twenty estimated $ATT$s was used for the body of this manuscript. We observe here that the empirical standard deviation of the estimates closely resembles the theoretically derived Abadie-Imbens standard error, which explicitly accounts for the uncertainty of the matching procedure.

\section{Treated and Control Franchise Frequency}
\label{sec:tcfreq}

\begin{figure}[hbtp!]
    \centering
    \includegraphics[scale=0.8]{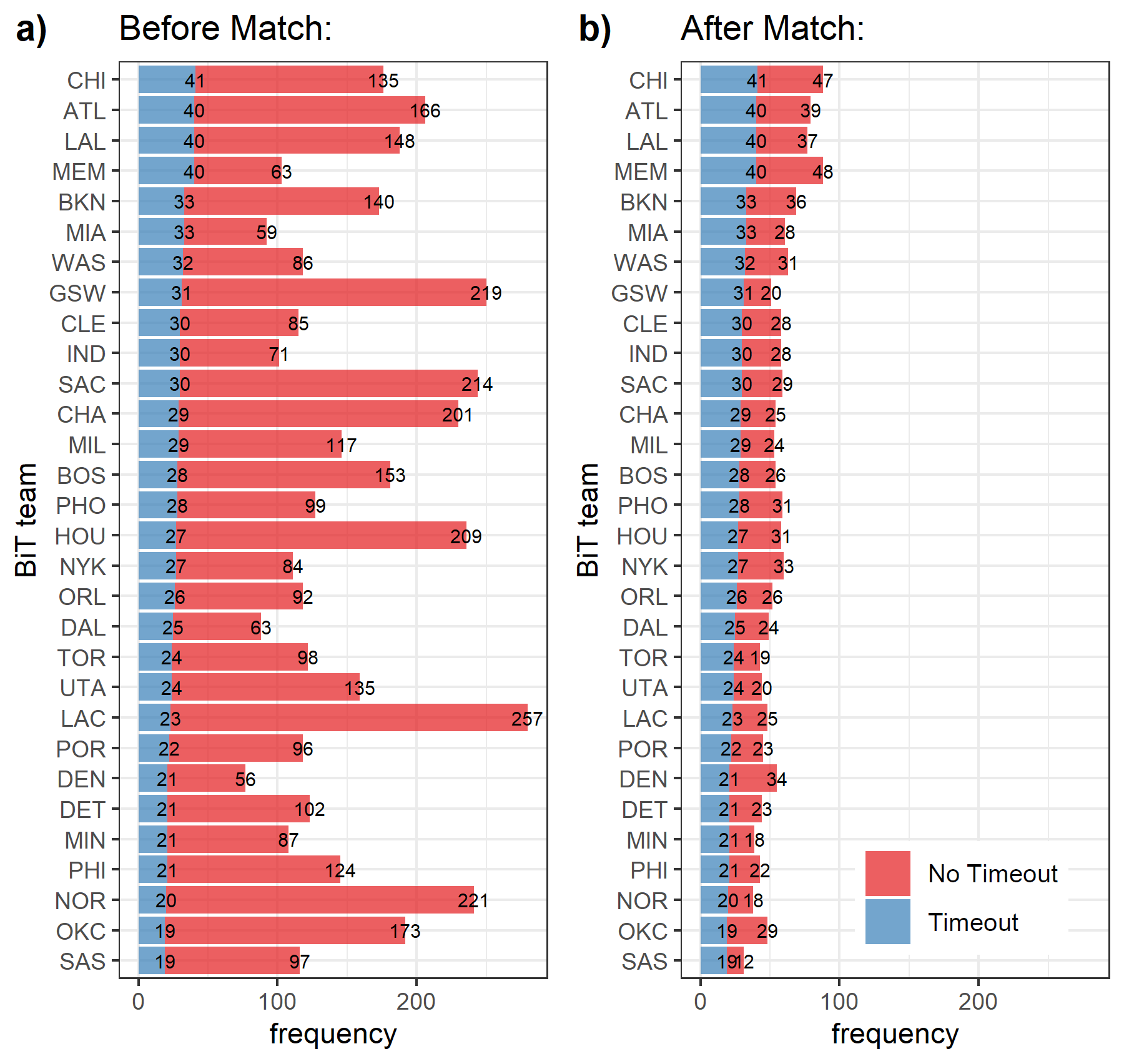}
    \caption{The blue horizontal bars indicate the number of runs with a timeout (RwTs) for a given franchise. The red horizontal bars indicate the number of runs without a timeout (RwoTs) for a given franchise. Each of these runs adhere to the criteria and are thus considered a unit. The matched set allows for estimation of the causal estimand after invoking the assumptions therein. There is no evidence of issues with positivity in estimating franchise specific effects since all teams exhibit instances of runs with and without timeouts.}
    \label{fig:franmat2}
\end{figure}

To investigate each franchise's strategy in responding to an opposing run (\emph{i.e.}, as the BiT team), we count the number of runs with and without a timeout before and after matching for each franchise, provided in Figure \ref{fig:franmat2}. We find that no franchise adheres to one and only one strategy when faced with an opposing run. In fact, the teams with the fewest recorded number of runs with a timeout called a timeout at only roughly half the frequency as the team with the most recorded number of runs with a timeout, 19 versus 41, respectively. Since no franchise adheres to one and only one strategy, implying any franchise can viably serve as a matched control, positivity is likely not an issue in estimation of the franchise effects.

Furthermore, since each franchise called a timeout during an opposing run at least 19 times, the franchise-specific effects of a timeout can be estimated; however, some franchises suffer from relatively small sample sizes when compared to other franchises. We take this into consideration by reporting the standard errors alongside the point estimates for the team effects in Figure 8 of the manuscript. The standard errors, as expected, ingrain the inequity in the number of observations by franchise where teams with more observations (Chicago Bulls) have much smaller standard errors than those teams with fewer observations (San Antonio Spurs).

\begin{figure}[hbtp!]
    \centering
    \includegraphics[scale=0.9]{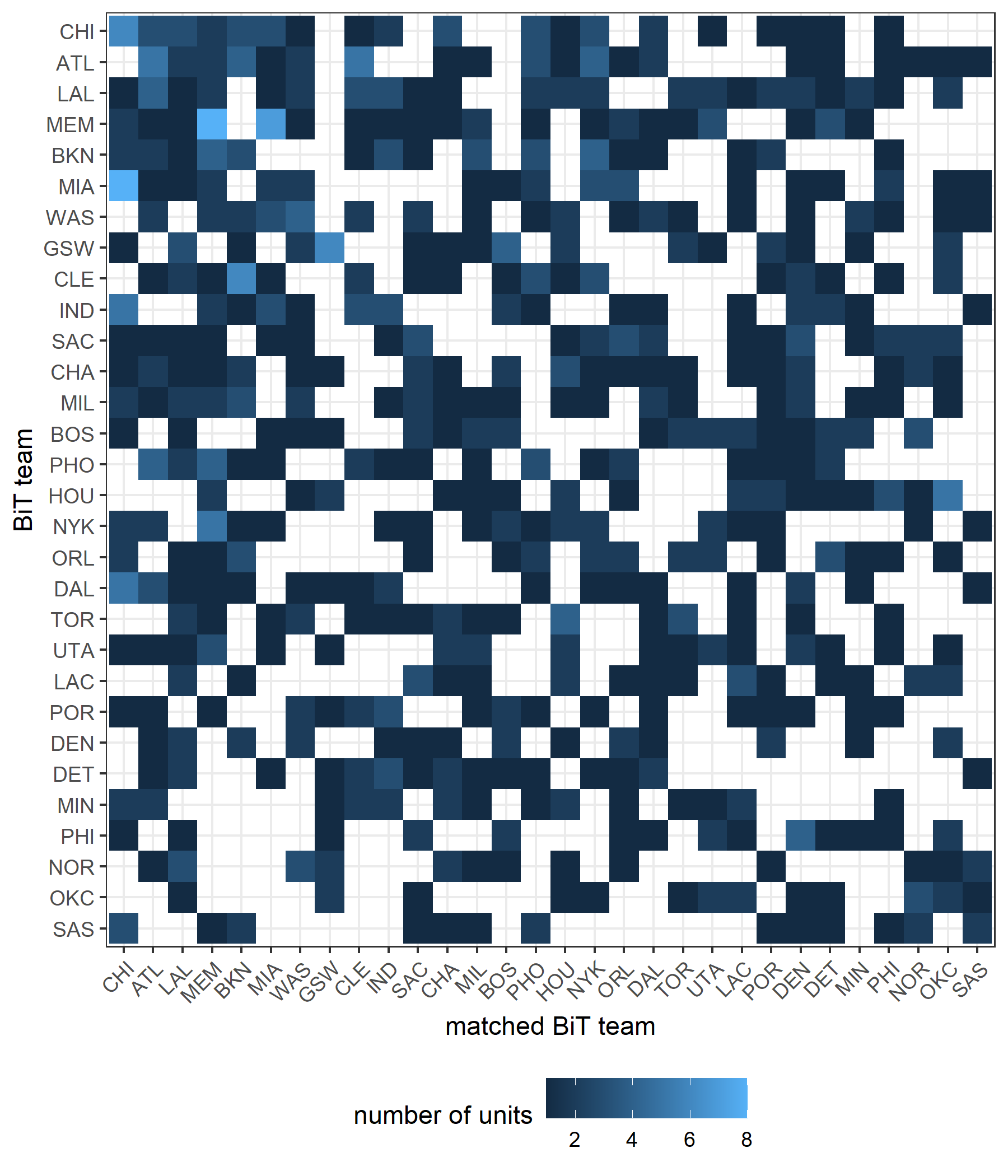}
    \caption{The number of matched controls by franchise for each franchise treated unit. The $(1,2)$ fill of the tile plot, for example, illustrates the number of times that an Atlanta Hawks run without a timeout was matched to a Chicago Bulls run with a timeout. No franchise is matched to a concernedly small number of franchises. The rows and columns of the tile plot have been arranged by the frequency with which that franchise serves as a matched control. That is, the New Orleans Pelicans were most often matched as a control (53), whereas the Denver Nuggets, Phoenix Suns, and Toronto Raptors were tied for least often matched as a control (14).}
    \label{fig:match-counts}
\end{figure}

After matching, we tabulate instances for which each franchise serves as the treatment (BiT team with a timeout) and matched control (BiT team without a timeout) in Figure \ref{fig:match-counts}. From this figure, we conclude that no franchise is matched to a concernedly small number of franchises. The sparsity of the plot is largely a function of the small sample size for several franchises, considering that slightly more than half of the franchises (19) have fewer runs with timeouts than the number of franchises (30) in the National Basketball Association (NBA).

\section{Covariate Balance}

\begin{figure}[hbtp!]
    \centering
    \includegraphics[scale=0.49]{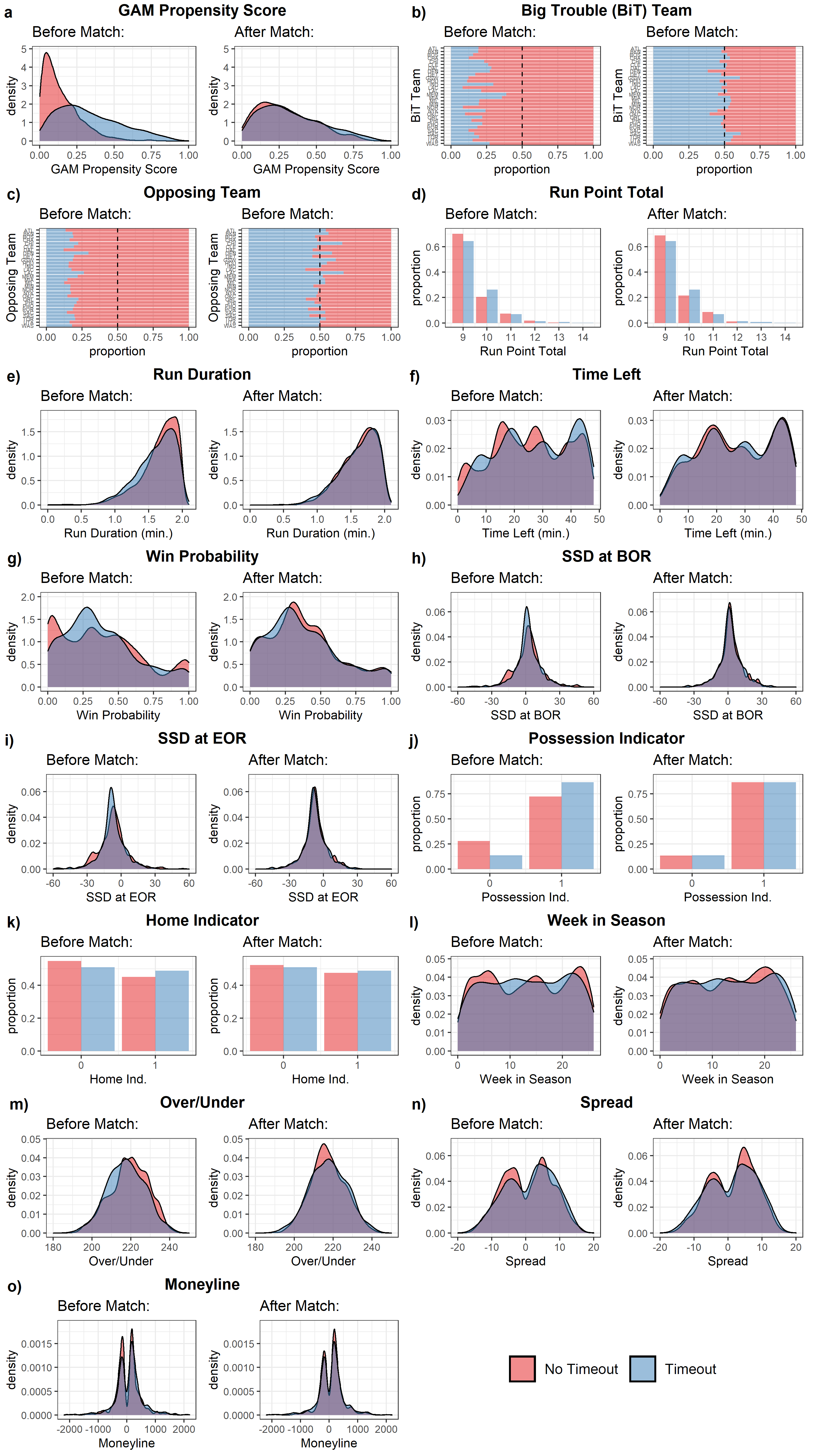}
    \caption{Distribution of the covariates for each treatment group before and after matching. Visually, matching appears to yield distributions of covariates which are similar across treatment group.}
    \label{fig:covariate-balance}
\end{figure}

To assess covariate balance after matching, distributions of the covariates for the treated and control units were created and shown in Figure \ref{fig:covariate-balance}. Before matching, the distribution of the covariates across treatment groups appear relatively similar for most covariates; however, there appear to be discrepancies in the BiT team, opposing team, win probability, time left, and possession across treatment group. Most notably, the distribution of the estimated propensity scores between the treatment groups was largely different. Aside from covariate imbalance, there appears to be a potential violation of positivity in the moneyline. After removing plays with a moneyline larger than $2,400$ in absolute value and invoking the matching procedure, the distribution of the estimated propensity scores appears largely balanced. Many of the discrepancies in the distributions of the covariates noted earlier seem to no longer exist, such as that in the win probability. Formal hypothesis testing was applied to each of the covariates to verify these visual interpretations. 

\setcounter{table}{2}

\begin{table}
\caption{Unadjusted $p$-values from hypothesis testing to assess for a discrepancy in the distribution of covariates before and after matching. Before matching, every covariate except for week in season exhibits imbalance between treatment groups. After matching, every covariate exhibits balance between treatment groups. In each case, multiple comparison correction is preformed to control the false discovery rate at 0.05, according to \citet{benjamini1995controlling}.}
\centering
\begin{tabular}[htbp!]{lrr}
\toprule
\multicolumn{1}{c}{} & \multicolumn{2}{c}{$p$-value} \\
\cmidrule(l{3pt}r{3pt}){2-3}
 & Pre-Match & Post-Match\\
\midrule
GAM Propensity Score & $<0.001$ & 0.038\\
Big Trouble (BiT) Team & $<0.001$ & 0.989\\
Opposing Team & 0.046 & 0.568\\
Run Point Total & 0.002 & 0.082\\
Run Duration & $<0.001$ & 0.842\\
\addlinespace
Time Left & $<0.001$ & 0.502\\
Win Probability & $<0.001$ & 0.064\\
SSD at BOR & $<0.001$ & 0.176\\
SSD at EOR & 0.002 & 0.138\\
Possession Indicator & $<0.001$ & 0.912\\
\addlinespace
Home Indicator & 0.037 & 0.619\\
Week in Season & 0.297 & 0.493\\
Over/Under & $<0.001$ & 0.145\\
Spread & 0.003 & 0.374\\
Moneyline & 0.004 & 0.112\\
\bottomrule
\end{tabular}
\label{tbl:covbal-p}
\end{table}
 
 To check for discrepancies in the distributions of discrete and continuous covariates across treatment groups, bootstrapped Kolmogorov-Smirnov tests were applied. For binary variables, $t$-tests were used, and chi-squared tests were used for categorical variables. The raw $p$-values for each of these tests before and after matching are provided in Table \ref{tbl:covbal-p}. Before matching, a discrepancy between the distributions of covariates associated with treatments and controls was identified in all covariates and the estimated propensity scores except for week in season which appeared sufficiently balanced. After matching, there was no evidence of discrepancy in covariate distributions for any of the covariates or the estimated propensity score. All covariates appear sufficiently balanced after matching. In each case, multiple comparison correction was preformed to control the false discovery rate at 0.05, according to \citet{benjamini1995controlling}.

\begin{figure}[hbtp!]
    \centering
    \includegraphics[scale=0.85]{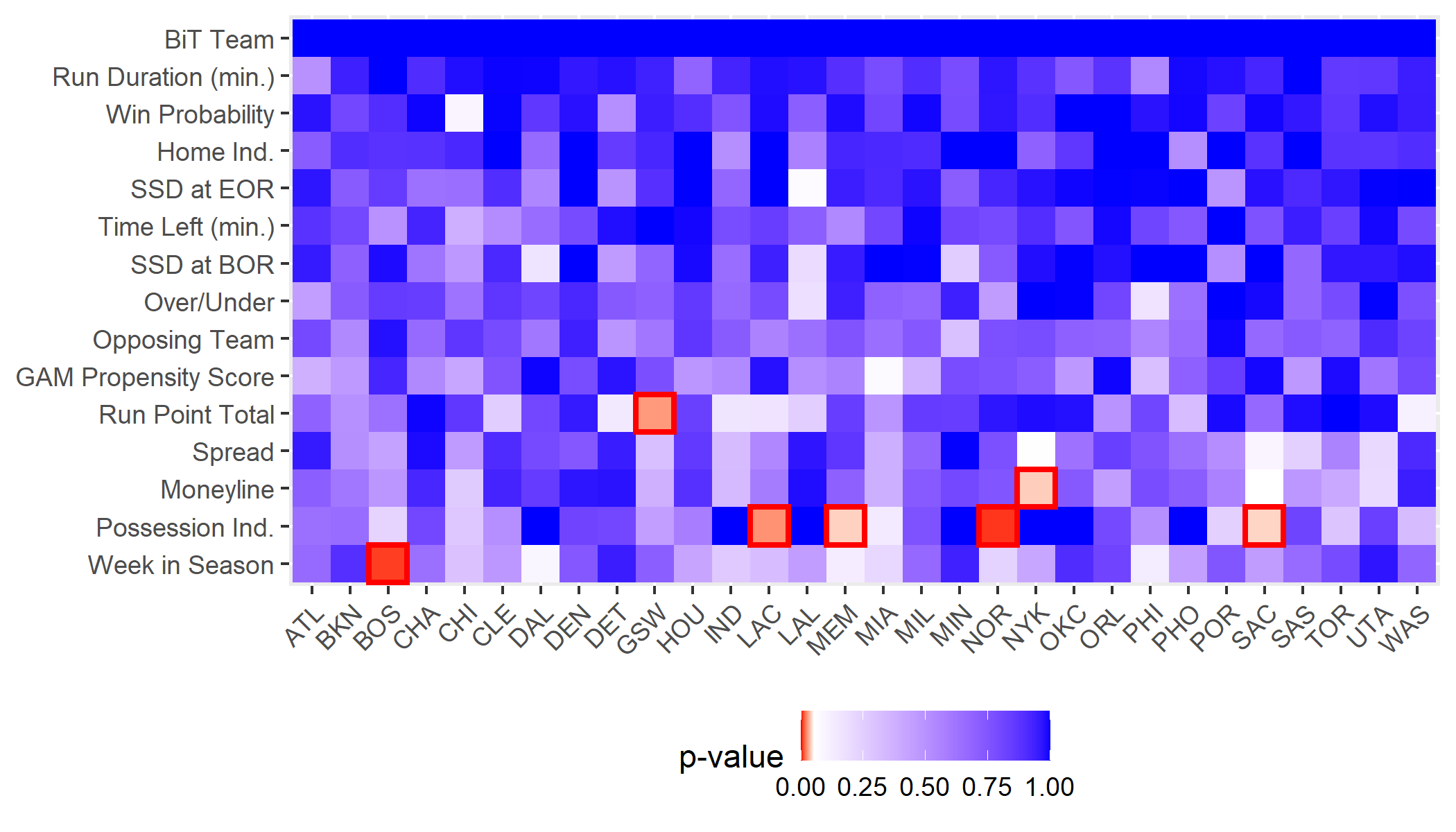}
    \caption{Hypothesis testing was conducted on the matched samples' covariates for each franchise. Unadjusted $p$-values for each covariate/franchise are shown. Seven p-values (outlined in red) maintained an unadjusted $p$-value less than 0.05, the minimum of which was $p=0.004$, corresponding to possession with the New Orleans Pelicans. After multiple comparison correction to control the false discovery rate at 0.05, no $p$-values are statistically significant. Hence, it is reasonable to assume the covariates are sufficiently balanced when conditioning on the BiT team's identity.}
    \label{fig:cfranignor}
\end{figure}

To assess covariate balance when conditioning on franchise, we perform hypothesis testing to assess potential discrepancies in the distribution of covariates between the treatment units (runs with a timeout) and control units (runs without a timeout) for each of the fifteen covariates and each of the thirty franchises. In all, 450 hypothesis tests were performed (15 covariates by 30 franchises), so a multiple comparison correction was performed to control the false discovery rate at 0.05. We attain an unadjusted $p$-value for each covariate/franchise combination (see Figure \ref{fig:cfranignor}). There are seven covariate/franchise combinations with an unadjusted $p$-value less than 0.05 (outlined in red in Figure \ref{fig:cfranignor}) with the smallest being $p=0.004$, corresponding to possession with the New Orleans Pelicans. After invoking the multiple comparison correction \citep{benjamini1995controlling}, there is no evidence for distributional discrepancy between matched sets of covariates for any of the franchises. This suggests it is reasonable to assume covariate balance when conditioning on the treated team's identity (i.e. the BiT team's identity) and to proceed with estimating the franchises' respective causal effects.

\section{Data and Code}

Data and code to reproduce the results in the paper are available in the following public GitHub repository: \url{https://github.com/ConGibbs10/nba-causal}.
\end{appendices}

\end{document}